\documentclass[final,3p,times,twocolumn]{elsarticle}

\usepackage{lineno}
\usepackage{graphicx}
\usepackage[colorinlistoftodos]{todonotes}
\usepackage{amssymb}
\usepackage{braket}

\usepackage{xspace}

\biboptions{comma,sort&compress}

\usepackage{amsmath}
\usepackage{siunitx}
\usepackage{cleveref}
\usepackage{CJKutf8}
\usepackage{rotating}
\usepackage{booktabs}
\usepackage{subfig}
\usepackage[version=4]{mhchem}
\usepackage{mathtools, cuted}

\journal{Journal of Molecular Spectroscopy}

\usepackage{xcolor,textcomp}

\begin{document}

\begin{frontmatter}


\title{(Sub-)millimeter-wave spectroscopy of \textit{gauche}-propanal}


\author[ph1]{Oliver Zingsheim\corref{cor1}}
\ead{zingsheim@ph1.uni-koeln.de}
\author[ph1]{Holger S. P. M\"uller\corref{cor1}}
\ead{hspm@ph1.uni-koeln.de}
\author[ph1]{Luis Bonah}
\author[ph1]{Frank Lewen}
\author[ph1]{Sven Thorwirth}
\author[ph1]{Stephan Schlemmer}
\cortext[cor1]{Corresponding author}


\address[ph1]{I. Physikalisches Institut, Universit\"at zu K\"oln, Z\"ulpicher Stra{\ss}e 77, 50937 K\"oln, Germany}

\begin{abstract}
A detailed analysis of (sub-)millimeter-wave spectra of the vibrational ground state ($\upsilon=0$) combined with the  energetically lowest excited vibrational state ($\upsilon_{24}=1$; aldehyde torsion) of \textit{gauche}-propanal (\textit{g}-\ce{C2H5CHO}) up to 500\,GHz is presented.
Both vibrational states, $\upsilon=0$ and $\upsilon_{24}=1$, are treated with tunneling rotation interactions between their two respective tunneling states, which originate from two stable degenerate \textit{gauche}-conformers; left- and right-handed configurations separated by a small potential barrier.
Thanks to double-modulation double-resonance (DM-DR) measurements, important but weak $c$-type transitions connecting the tunneling states could be unambiguously assigned.
In addition, Coriolis interaction as well as Fermi resonance between the two vibrational states needed to be taken into account to derive fits with experimental accuracy using Pickett's SPFIT program in a reduced axis system (RAS).
Based on the rotational analysis, the fundamental vibrational frequency $\nu_{24}$ of \textit{gauche}-propanal is redetermined to 68.75037(30)\,cm$^{-1}$.
\end{abstract}

\end{frontmatter}


\section{Introduction}                        
\label{Sec:Introduction}                      

\begin{figure}[t!]
\centering
\includegraphics[width=1.0\linewidth]{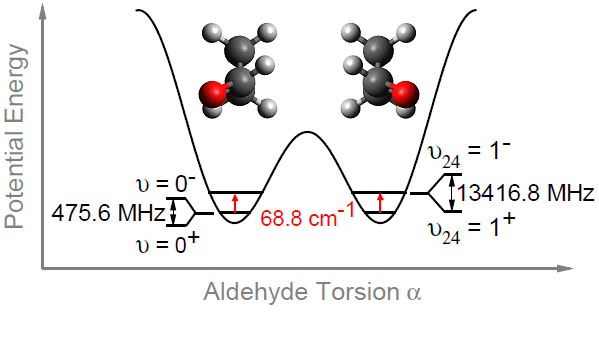}
\caption{Schematic double-well potential function of the aldehyde rotation for \textit{gauche}-propanal. The given energy splittings of the tunneling states ($\upsilon=0^+$ and $\upsilon=0^-$ as well as $\upsilon_{24}=1^+$ and $\upsilon_{24}=1^-$) and the fundamental frequency $\nu_{24}$ (in red; $\upsilon_{24}=1\leftarrow 0$) are the main results of this article.}
\label{Fig:1_potential}
\end{figure}

Rotational and vibrational spectroscopic fingerprints of propanal, or propionaldehyde (\ce{CH3CH2CHO}), were subject of several studies \cite{Propanal_1_Butcher_Wilson,Propanal_2_Pickett_Scroggin,Propanal_3_JAH,Propanal_4_Randell_Gauche,Propanal_5_Demaison,Propanal_6_Structure,Propanal_7_PotentialFunction,Propanal_8_FIR,Propanal_9_IR,ZINGSHEIM2017_Propanal}. \citet{Propanal_1_Butcher_Wilson} have proven the existence of two stable conformers, \textit{syn}- and \textit{gauche}-propanal ($C_s$ and $C_1$ symmetry, respectively). These differ mainly in the rotation of the aldehyde group ($-$CHO).
Another large amplitude motion (LAM) is the methyl group ($-$CH$_3$) internal rotation; its barrier height $V_3$ is 793.7(25)\,cm$^{-1}$ \cite{Propanal_3_JAH} or 784.2(90)\,cm$^{-1}$ \cite{Propanal_5_Demaison} for \textit{syn}-propanal, which was determined from microwave and millimeterwave studies, respectively. 
\textit{Gauche}-propanal has a comparable barrier height of $V_3=886(10)$\,cm$^{-1}$ \cite{Propanal_4_Randell_Gauche}.
Furthermore, vibrational modes of propanal were assigned by low-resolution far-infrared (FIR) \cite{Propanal_8_FIR} and mid-IR studies \cite{Propanal_9_IR}.
The energetically lowest lying fundamental frequency, the aldehyde torsion $\upsilon_{24}=1\leftarrow0$, of \textit{syn}- and \textit{gauche}-propanal was determined to be $\nu_{24}=135.1$\,cm$^{-1}$ and 113.1\,cm$^{-1}$, respectively \cite{Propanal_8_FIR}.
The later value will be redetermined with the rotational analysis presented here.
Tunneling between the two degenerate \textit{gauche} configurations is feasible and leads to symmetric ($\upsilon^+$) and antisymmetric ($\upsilon^-$) tunneling states of each vibrational state due to a small potential barrier separating the \mbox{"left-"} and "right-handed" \textit{gauche} configurations (The tunneling transition state has $C_s$ symmetry) \cite{Propanal_1_Butcher_Wilson}.
One main aim of this article is to study the tunneling process of the aldehyde-group in \textit{gauche}-propanal, in particular to determine accurate energy splittings of $\upsilon=0$ and $\upsilon_{24}=1$, see Fig.~\ref{Fig:1_potential}.
These splittings together with accurate vibrational frequencies of $\upsilon_{24}$ for \textit{syn}- and \textit{gauche}-propanal may allow to more precisely model the potential describing the interconversion of both conformers in the future.

Studying the two LAMs of propanal is of fundamental interest itself, likewise its astronomical relevance.
Therefore, another aim of this study is to warrant a solid quantum mechanical model of \textit{gauche}-propanal to facilitate astronomical detection.
\textit{Syn}-propanal was already detected in space \cite{First_Detection_Propanal,Requena_Torres_2008,Lykke_space}, in contrast to \textit{gauche}-propanal.
The detection of \textit{syn}-propanal in comparably warm environments of around 125\,K \cite{Lykke_space} may give some confidence for a future detection of its higher energy conformer.
Unfortunately, the description of the vibrational ground state of \textit{gauche}-propanal is somewhat limited as observable interactions with energetically higher excited vibrational states have been neglected so far \cite{ZINGSHEIM2017_Propanal}. The previoulsy determined fundamental frequency $\nu_{24}$ \cite{Propanal_8_FIR,Propanal_9_IR} was too large to account for interactions between $\upsilon=0$ and $\upsilon_{24}=1$.
Notable deviations from a single state fit of $\upsilon=0$ with 2378 transitions were observed for several $R$-branch $a$-type transitions ($^qR$ series) \cite{ZINGSHEIM2017_Propanal}, even though tunneling-rotation interaction between the two tunneling states of $\upsilon=0$, hereafter referred to as $0^+$ and $0^-$, were taken into account up to 500\,GHz \cite{Propanal_2_Pickett_Scroggin,ZINGSHEIM2017_Propanal}.
On the other hand, tunneling-rotation interaction were not taken into account for $\upsilon_{24}=1$ and only 15 $a$-type rotational transitions with $J\leq4$ and $K_a\leq2$ were assigned but not fit to experimental accuracy \cite{Propanal_7_PotentialFunction}; the tunneling states of the aldehyde torsion ($\upsilon_{24}=1$) are hereafter abbreviated with $24^+$ and $24^-$. Therefore, a combined analysis of $\upsilon=0$ and $\upsilon_{24}=1$ is needed.

A detailed analysis of the fingerprints of \textit{gauche}-propanal is presented in this article. The experimental setups measuring these fingerprints are summarized next (Sec.~\ref{Sec:Experiment_Setup}). Conventional 2$f$ modulated broadband measurements and double-modulation double-resonance (DM-DR) measurements were performed.
DM-DR measurements allow unambiguous assignments of weak or perturbed transitions in a straightforward fashion, as its DR character finds linkages, therefore warrants the correct assignment, and the DM character clears up crowded spectra \cite{Zingsheim2021_DMDR}. 
The analysis of the rotational fingerprint of \textit{gauche}-propanal is described in detail (Sec.~\ref{Sec:Rot_Fingerprints}). Theoretical considerations, in particular the applied Hamiltonian, and a first single state analysis of $\upsilon_{24}=1$ are presented to illustrate characteristics of the observed spectroscopic fingerprints. The section is completed with a combined analysis of $\upsilon=0$ and $\upsilon_{24}=1$ which takes Fermi resonance and Coriolis interaction into account.
Finally, the main results and future prospects are discussed (Sec.~\ref{Sec:Discussion}).

\bigskip

\section{Experimental Setup}                     
\label{Sec:Experiment_Setup}                     

Conventional absorption measurements up to 500\,GHz were already performed and their experimental conditions were published in Ref.~\cite{ZINGSHEIM2017_Propanal}.
Some additional measurements in the W-band were performed for this study, both conventional and DM-DR ones, to increase the accuracy of center frequencies as well as to search for linkages and to assign weak $c$-type transitions, respectively. 
In the following, only brief summaries of the experimental setups are presented as both, i) the conventional \cite{Drumel2015_OSSO,Ordu2019_Acetone} and the ii) DM-DR setup \cite{Zingsheim2021_DMDR}, were already described in detail before.

i) The probe signal of a synthesizer (up to 43\,GHz) is multiplied by different (cascaded) multipliers to reach a frequency coverage of 75$-$500\,GHz for conventional measurements. Some additional measurements below 67\,GHz are done without multiplication by straightforwardly using the output frequency of an Agilent E8257D synthesizer.
The frequency modulated radiation is guided with lenses and reflected by a rooftop mirror through the absorption glass cell containing solely the molecules of interest with pressures around 20\,\textmu bar. The double-pass setup has an absorption path length of $2\times5$\,m with a cell diameter of 10\,cm \cite{Drumel2015_OSSO}.
A polarization grid is guiding the radiation onto a Schottky detector (and can separate two beams in DR measurements), then the signal is amplified and processed by a bandpass-filter. Finally, the 2$f$ demodulation of a lock-in amplifier ($f\approx47$\,kHz) makes absorption features appear close to the second derivative of a Voigt-profile.

ii) A second synthesizer and a 14\,m single pass setup are used for DM-DR measurements in the W-band region \cite{Zingsheim2021_DMDR}. In this setup, the pump radiation with output powers of up to 60\,mW (AFM6 70–110+14 from RPG) can be superimposed with the probe radiation ($\approx1$\,mW), orthogonally polarized to each other.
By additional modulation of the pump source, a difference spectrum of DR on and DR off measurements is created, which contains only signals if probe and pump frequencies are resonant to two transitions which share an energy level, mainly based on the Autler-Townes effect \cite{Autler_Tones_1955,Tannoudji_DressedAtomApproach}.
This method results in confusion- and baseline-free measurements allowing to unambiguously assign weak and perturbed transitions in a straightforward fashion \cite{Zingsheim2021_DMDR}.

\section{Rotational Fingerprint}                
\label{Sec:Rot_Fingerprints}                    

Specific molecular properties and theoretical considerations which describe the rotational fingerprint of \textit{gauche}-propanal are summarized first (Sec.~\ref{SubSec:Theory}).
Based on these considerations, a single vibrational state analysis of $\upsilon_{24}=1$ is discussed (Sec.~\ref{SubSec:v24}).
This single state analysis turned out to have similar limitations as the recent analysis of $\upsilon=0$ \cite{ZINGSHEIM2017_Propanal}.
Deviations of adjacent $^qR$ series transitions with specific $K_a$ quantum numbers ($K_a=12$ and $K_a=14$ for $\upsilon=0$; $K_a=7$ and $K_a=10$ for $\upsilon_{24}=1$) can be circumvented by a combined analysis of the two vibrational states, or strictly speaking of the four tunneling states (Sec.~\ref{SubSec:v0+v24}).

\subsection{Theoretical considerations}       
\label{SubSec:Theory}                         

\textit{Gauche}-propanal is quite close to the limiting prolate rotor case with $\kappa=-0.9849$.
The dipole moment components of \textit{gauche}-propanal were determined to be $\mu_a=2.645(5)$\,D, $\mu_b=0.417(6)$\,D, and $\mu_c=1.016(3)$\,D \cite{Propanal_7_PotentialFunction}.
As the symmetry plane of the \textit{gauche}-propanal tunneling transition state is coinciding with the $ab$-plane of the molecule, parity selection rules make $a$- and $b$-type transitions occur within tunneling states ($\upsilon^+\leftrightarrow \upsilon^+$ or $\upsilon^-\leftrightarrow \upsilon^-$), whereas $c$-type transitions occur between the two states ($\upsilon^+\leftrightarrow \upsilon^-$). For this reason, $c$-type transitions contain direct information about the energy difference $\Delta E_\upsilon$ of the tunneling states.

In general, the Hamiltonian 
\begin{equation}
\hat{H}=
\begin{bmatrix}\hat{H}_{\nu''} & \hat{H}_{\nu'\nu''} \\
\hat{H}_{\nu'\nu''} & \hat{H}_{\nu'}+\Delta E \end{bmatrix} \\
\label{Eq:Hamiltonian_interaction}
\end{equation}
used to describe the interaction between two states
consists of the common asymmetric rotor Hamiltonians on the diagonal, one for each of the two distinct vibrational states, $\hat{H}_{\nu'}$ and $\hat{H}_{\nu''}$, their energy difference $\Delta E$, and their interaction $\hat{H}_{\nu'\nu''}$.
The off-diagonal elements may take Fermi resonance and Coriolis interaction into account ($\hat{H}_{\nu'\nu''}=\hat{H}^F_{\nu'\nu''}+\hat{H}^C_{\nu'\nu''}$).

Fermi parameters used in the analysis of \textit{gauche}-propanal are of the form \cite{dyad_F_plus_minus,KISIEL200926,muller_cl_coriolis} 
\begin{equation}
\hat{H}^F_{\nu'\nu''} = F_0 + F_J \hat{J}^2 + F_K \hat{J_a}^2 + 2 F_2 (\hat{J_b}^2-\hat{J_c}^2) +...
\end{equation}
and the Coriolis terms are of the form
\begin{equation} 
\begin{split}
\hat{H}^C_{\nu'\nu''}=
+~&iG_a\hat{J}_a + iG_b\hat{J}_b + iG_c\hat{J}_c ~~~~~~~~~~~~~~~~~~~~~~~~~~~(n=1) 
\\
+~&...\\
+~&F_{bc}(\hat{J}_b\hat{J}_c+\hat{J}_c\hat{J}_b) + F_{ca} (...) + F_{ab}(...)~~~~~(n=2)
\\
+~&F_{bc}^J\hat{J}^2(\hat{J}_b\hat{J}_c+\hat{J}_c\hat{J}_b)
+F_{bc}^K\{(\hat{J}_b\hat{J}_c+\hat{J}_c\hat{J}_b),\hat{J}^2_a\}/2
\\
+~&...
\end{split}
\label{Eq:Coriolis_Terms_specific}
\end{equation}
with $\{,\}$ being the anticommutator.
Coriolis parameters and their respective operators are also given in Tables 3 and 4 of Ref.~\cite{Drouin_Spfit_overview}.

In total, the rotational Hamiltonian of the combined analysis of $\upsilon= 0$ and $\upsilon_{24}= 1$ is described in the basis of the four tunneling states $(\psi_{0^+},\psi_{0^-},\psi_{24^+},\psi_{24^-})$.
It is explicitly described by Eq.~\eqref{Eq:Hamiltonian_Full}.

An iterative fitting procedure of assigning, fitting and predicting spectra is applied. 
The assigned transitions are fit with Pickett's SPFIT program \cite{PICKETT1991} using a reduced axis system (RAS) to treat molecules affected by tunneling-rotation interaction, which Pickett proposed in 1972 \cite{RAS_Coriolis}.
The RAS was successfully adapted in the description of many molecules, e.g. in Refs. \cite{cyanamide_rot_1986,cyanamide_isos_rot_2011,cyanamide_w_15N_13C_rot_2019,ethanediol_rot_2003,ethanediol_rot_2004,ethanediol_rot_2020,propargyl__alcohol_rot_2005,H2DO+_analysis_rot_2010,ethanethiol_RAS-Fit_2016,hydroxyacetonitrile_rot_2017,hydroxymethyl_rot_2017,hydroxymethyl_rot_2020}.
Furthermore, it also allowed to describe strongly perturbed transitions of $\upsilon= 0$ of \textit{gauche}-propanal \cite{Propanal_2_Pickett_Scroggin,ZINGSHEIM2017_Propanal}.
Pickett's RAS is also described by Coriolis parameters, but only operators of even order $n$ are used \cite{RAS_Coriolis}.
The Hamiltonians $\hat{H}_{X^\pm}$ in Eq.~\ref{Eq:Hamiltonian_Full} account for tunneling rotation interaction in the RAS.
In general, if the species of rotation $R_g$ is
contained in the direct product of the symmetry species $g$ (with $g = a; b; c$), Coriolis interactions can occur between two states \cite{muller_cl_coriolis}.
For \textit{gauche}-propanal with the tunneling transition state being of $C_s$ symmetry, $F_{ab}$ ($c$-type)
is acting within each tunneling state ($A' \oplus 
 A' = A'$), in contrast to
$F_{bc}$ and $F_{ca}$  ($a$- and $b$-type interactions are expected to occur between the two substates as $A' \oplus A'' = A''$).
Therefore, $a$- and $b$-type tunneling-rotation interaction, $\hat{H}^{a/b}_{X^\pm}$, are off-diagonal in Eq.~\eqref{Eq:Hamiltonian_Full}, whereas the $c$-type ones, $\hat{H}^c_{X^\pm}$, are on the diagonal.
\newpage
\begin{strip}
\begin{equation}
\hat{H}=
\begin{bmatrix}
\hat{H}_{0^+}+\hat{H}^c_{0^\pm}-\Delta E_0/2 & \hat{H}^{a/b}_{0^\pm} & \hat{H}^F_+ & \hat{H}^C \\
\hat{H}^{a/b}_{0^\pm} & \hat{H}_{0^-}+\hat{H}^c_{0^\pm}+\Delta E_0/2 & \hat{H}^C& \hat{H}^F_-\\
\hat{H}^F_+ & \hat{H}^C& \hat{H}_{24^+}+\hat{H}^c_{24^\pm} +E_{24}- \Delta E_{24}/2 & \hat{H}^{a/b}_{24^\pm}\\
\hat{H}^C & \hat{H}^F_-& \hat{H}^{a/b}_{24^\pm} &  \hat{H}_{24^-}+\hat{H}^c_{24^\pm}  +E_{24}+ \Delta E_{24}/2 
\end{bmatrix}
\label{Eq:Hamiltonian_Full}
\end{equation}
\end{strip}

\newpage

\subsection{Single vibrational state analysis: $\upsilon_{24}=1$} 
\label{SubSec:v24}                            

\begin{figure}[t]
\centering
\includegraphics[width=1.0\linewidth]{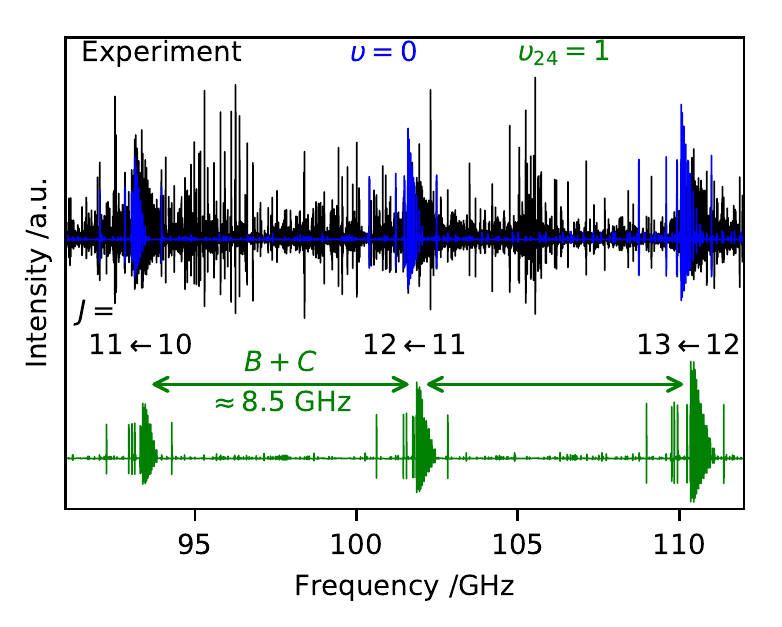}
\caption{Measured spectrum of propanal in the W-band region (in black).
In addition, predictions of the vibrational ground state ($\upsilon=0$; in blue) and of the energetically lowest vibrational excited state of \textit{gauche}-propanal ($\upsilon_{24}= 1$; in green) are shown.}
\label{Fig:2_gauche-satellites}
\end{figure}

The constantly repeated emergence of rather strong and unassigned lines next to predicted $K_a$ structures of $\upsilon=0$ of \textit{gauche}-propanal (blue) in Fig.~\ref{Fig:2_gauche-satellites} reveal the presence of a rather strong satellite spectrum belonging to $\upsilon_{24}=1$ (green).
$K_a$ structures are made of $^qR$ series transitions, $J_{K_a,K_c}\leftarrow (J-1)_{K_a,K_c-1}$,  with identical $J$ and are approximately separated by $B+C$ for adjacent $J$s for near prolate rotors, cf. Fig.~\ref{Fig:2_gauche-satellites}.

\begin{figure}[t]
\centering
\includegraphics[width=1.0\linewidth]{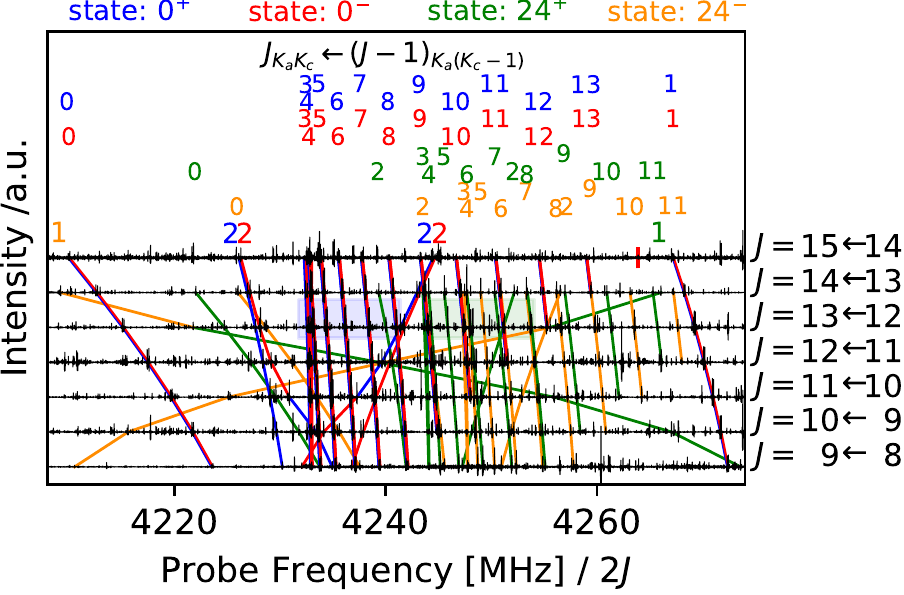}
\caption{Fortrat diagram of the conventional spectrum of \textit{gauche}-propanal in the W-band region (in black). $^qR$-series transitions, $J_{K_a,K_c}\leftarrow (J-1)_{K_a,K_c-1}$, with adjacent $J$s of $\upsilon_{24}=1$ were linked by DM-DR measurements and their connection are visualized by green and orange lines. Assignments of $\upsilon=0$ from an earlier study are shown in red and blue \cite{ZINGSHEIM2017_Propanal}. Transitions are prolate paired, except for $K_a\leq3$, and $K_a$'s of each $^qR$-series are given vertical above the last transition. The lowest row of $K_a$ quantum numbers represent strongly perturbed series (cross shape series) with $K_a=2$ of $\upsilon=0$ and $K_a=1$ of $\upsilon_{24}=1$ in the W-band region.}
\label{Fig:3_Fortrat-DR-links}
\end{figure}

The various $^qR$ series are often easily traceable by visual means, e.g., in Fortrat diagrams, see Fig.~\ref{Fig:3_Fortrat-DR-links}. 
From this point on a strict color coding of blue, red, green, and orange is used throughout the figures of this article to represent the tunneling states $0^+$, $0^-$, $24^+$, and $24^-$, respectively.

\begin{figure}[t]
\centering
\includegraphics[width=1.0\linewidth]{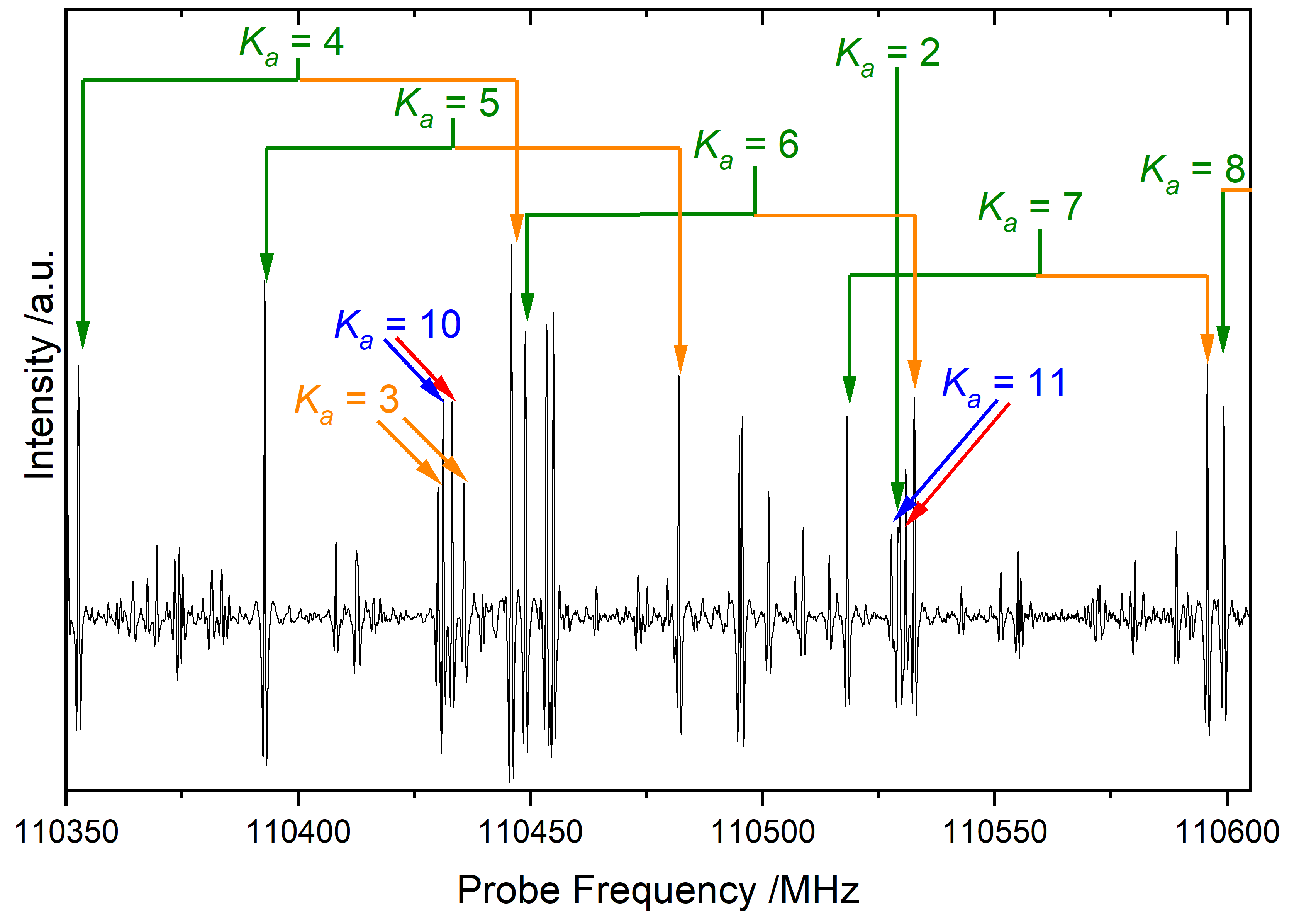}
\caption{Spectroscopic fingerprint of \textit{gauche}-propanal in the W-band region.}
\label{Fig:4_Fingerprint}
\end{figure}

Assigning $K_a$ and tunneling quantum numbers was less straightforward than in the $\upsilon=0$ case.
In particular the doublet pattern of $\upsilon_{24}=1$ was initially not as obvious as for $\upsilon=0$.
The DM-DR measurements greatly simplified the assignment procedure, as several $^qR$ series were experimentally linked, see green and orange lines in Fig.~\ref{Fig:3_Fortrat-DR-links}.
First of all, the assignment of $J$ quantum numbers to linked $^qR$ series transitions is straightforward thanks to the Fortrat diagram as $J\leftarrow (J-1)$ is known for each layer, see Fig.~\ref{Fig:3_Fortrat-DR-links}.
Without any additional \textit{ab-initio} calculations rather obvious peculiarities in Fig.~\ref{Fig:3_Fortrat-DR-links} helped to fully assign transitions in the W-band:\\
i) Transitions with $9\leq K_a \leq 11$ could be assigned because R-branch series start with $J=K_a$.
ii) Transitions with $K_a=0$ and $K_a=2$ were assigned subsequently as the slope of these linked transitions is rather different to other $^qR$ series, but comparable to those of $\upsilon=0$ with the same $K_a$'s, see Fig.~\ref{Fig:3_Fortrat-DR-links}.
iii) Transitions with $K_a=3$ are also recognizable because of their quartet structure being close in frequency as prolate pairing is slightly lifted.
iv) Assignments of transitions with $4\leq K_a \leq 8$ were then rather straightforward as they are located between $K_a=3$ and $K_a=9$ series.
v) Very importantly, the analysis profited furthermore immensely from DM-DR measurements as the unambiguous assignment of heavily perturbed transitions with $K_a=1$ was facilitated, see lines which form a big cross and change color from orange to green or vice versa in Fig.~\ref{Fig:3_Fortrat-DR-links}.
In the end, the doublets of $\upsilon_{24}=1$ are not as distinct as for $\upsilon=0$. Transitions of $24^+$ with $K_a+1$ appear sometimes at lower frequencies than transitions with $K_a$ of $24^-$ for identical $J$'s. A typical fingerprint of $\upsilon_{24}=1$ of \textit{gauche}-propanal is shown in Fig.~\ref{Fig:4_Fingerprint}.

\begin{figure}[t]
\centering
\includegraphics[width=0.9\linewidth]{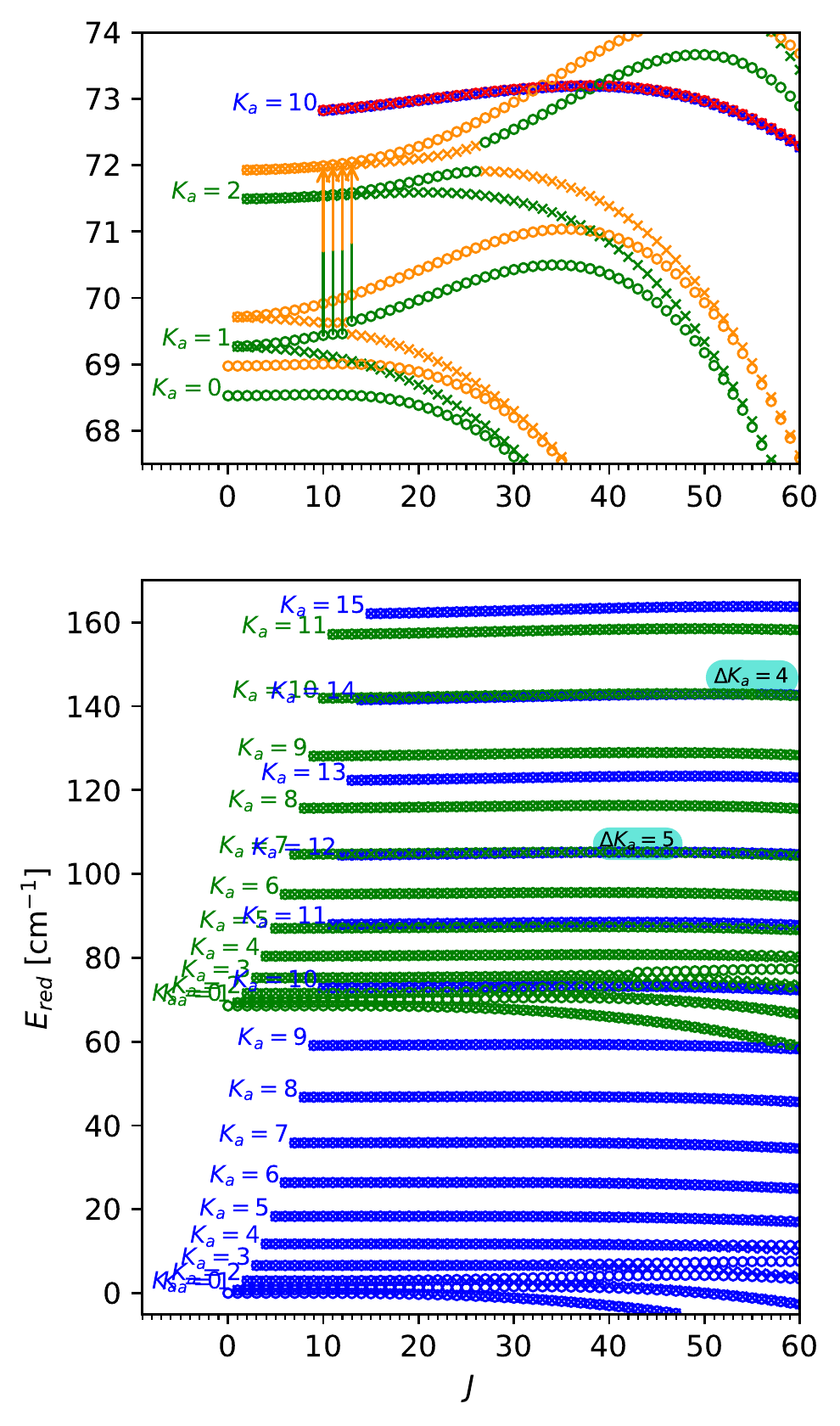}
\caption{
Reduced energy diagram for $\upsilon=0$ and $\upsilon_{24}=1$ of \textit{gauche}-propanal. In the bottom panel, only $0^+$ (blue) and $24^+$ (green) tunneling states are plotted for simplicity as tunneling states belonging to one vibrational state seem to overlap on this scale. In the top panel, energy levels of all states are plotted in an enlarged section, which highlights the $a$-type tunneling-rotation interaction ($F_{bc}$) of $24^+$ and $24^-$. Energy levels for asymmetry sides $K_a+K_c=J$ and $K_a+K_c=J+1$ are marked by circles ('o') and crosses ('x'), respectively, and are plotted with respect to their reduced energy $E_{red}= E - J(J + 1)(B + C)/2$.
}
\label{Fig:5_Re-overview}
\end{figure}

Assignments could be made up to 500\,GHz based on predictions derived from modeling the assigned transitions in the W-band region. 
The leading term for the strong $a$-type interaction, between energy levels with the same $K_a$ and $J$ values, observed in the W-band is $F_{bc}$, identical to the ground state analysis.
Basically, energy levels of the energetically higher asymmetry side $K_a+K_c=J$ of the lower tunneling state $24^+$ are interacting with the lower asymmetry side $K_a+K_c=J+1$ of the higher tunneling state $24^-$, see the top panel of the reduced energy diagram in Fig.~\ref{Fig:5_Re-overview}.
An avoided crossing pattern results for $K_a=1$, $K_a=2$, and $K_a=3$ around $J=12$, $J=26$, and $J=42$, respectively, clearly recognizable by trends of energy levels which switch from green circles to orange crosses or vice versa. This is also seen by the crossing series with $K_a=1$ in Fig.~\ref{Fig:3_Fortrat-DR-links}.

\begin{figure}[t]
\centering
\includegraphics[width=1.0\linewidth]{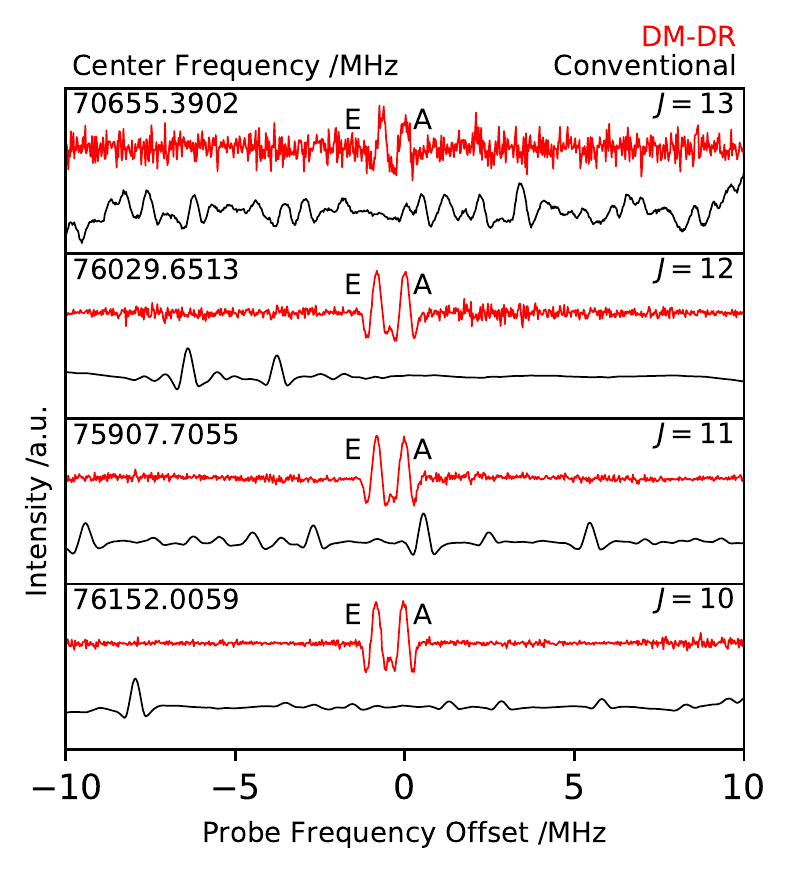}
\caption{DM-DR measurements (in red) of $Q$-branch $c$-type transitions. These transitions connect the two tunneling states ($J_{K_a,K_c}=J_{2,J-1}\leftarrow J_{1,J-1}$; $24^-\leftarrow24^+$) and are schematically depicted by the arrows in Fig.~\ref{Fig:5_Re-overview}. Assignments are hampered in conventional spectra (in black), which have been already mathematically processed (Fourier filtered) to remove the standing waves. $C$-type transitions show methyl internal rotation splitting (A, E).
}
\label{Fig:6_c-types}
\end{figure}

The analysis of $\upsilon_{24}=1$ was continued by assigning weak $c$-type transitions, again facilitated by DM-DR measurements. 
As already mentioned, $c$-type transitions directly connect the tunneling states and, therefore, play a key role in establishing the energy difference $\Delta E_{24}$ and with that also the interaction regions of the two tunneling states of $\upsilon_{24}=1$.
$C$-type transitions of strongly perturbed energy levels are of particular interest.
For example, four $Q$-branch $c$-type transitions $J_{K_a,K_c}=J_{2,J-1}\leftarrow J_{1,J-1}$ marked by the vertical arrows in Fig.~\ref{Fig:5_Re-overview} connect energy levels with $K_a=1$ of $24^+$ to levels with $K_a=2$ of $24^-$ and their DM-DR spectra are plotted in Fig.~\ref{Fig:6_c-types}. The avoided crossing behavior caused by the tunneling-rotation interactions is elucidated by the huge frequency difference of about 5374\,MHz between transitions with $J=12$ and $J=13$ (upper two panels of Fig.~\ref{Fig:6_c-types}).
Assigning these heavily perturbed transitions by using only conventional techniques would be very difficult.
$C$-type transitions of $\upsilon=0$ were assigned additionally, which was not possible in an earlier study \cite{ZINGSHEIM2017_Propanal}. Such an assignment of $\upsilon=0$ is given as a showcase example in the article presenting the newly developed DM-DR technique, see Fig.~8 in Ref.~\cite{Zingsheim2021_DMDR}.

$C$-type transitions are split measurably by the methyl group internal rotation in contrast to $a$-type transitions. 
A and E components can sometimes be experimentally distinguished if the asymmetry splitting is comparable to the internal rotation splitting because it leads to the presence of nominally forbidden E$^*$ transitions in rigid rotors which share their intensity with the E transitions \cite{Herschbach_PO_forbidden_E_lines}, demonstrated also with $Q$-branch $c$-type transitions $J_{K_a,K_c}=J_{3,J-2}\leftarrow J_{2,J-2}$ in the supplementary material (Fig.~\ref{FigA:3_c-types_unp}).
Only the A components are fit as only 85 out of 1977 assigned transitions of $\upsilon_{24}=1$ in the final analysis are $c$-type transitions, and these are too weak for a detection in space, hence a full treatment of the internal motion is neglected and A and E assignments should be re-examined. 

The single vibrational state analysis is able to reproduce large fractions of $^qR$ series transitions up to $K_a=11$ with $J_{max}=60$ of $\upsilon_{24}=1$, only limited by the frequency cutoff at 500\,GHz, and $c$-type transitions to experimental accuracy. However, $^qR$ series transitions with $K_a=7$ and $J>36$ as well as with $K_a=10$ and $J>30$ have been excluded here.
The deviations, $\nu_{\text{Obs.}}-\nu_{\text{Calc.}}$, of these perturbed transitions show patterns of an avoided crossing behavior originating from an interaction with another vibrational state ($\upsilon=0$).

\subsection{Combined analysis: $\upsilon=0$ and $\upsilon_{24}=1$}   
\label{SubSec:v0+v24}                         

\begin{figure}[t]
\centering
\begin{minipage}[r]{.225\textwidth}
  \includegraphics[width=\linewidth]{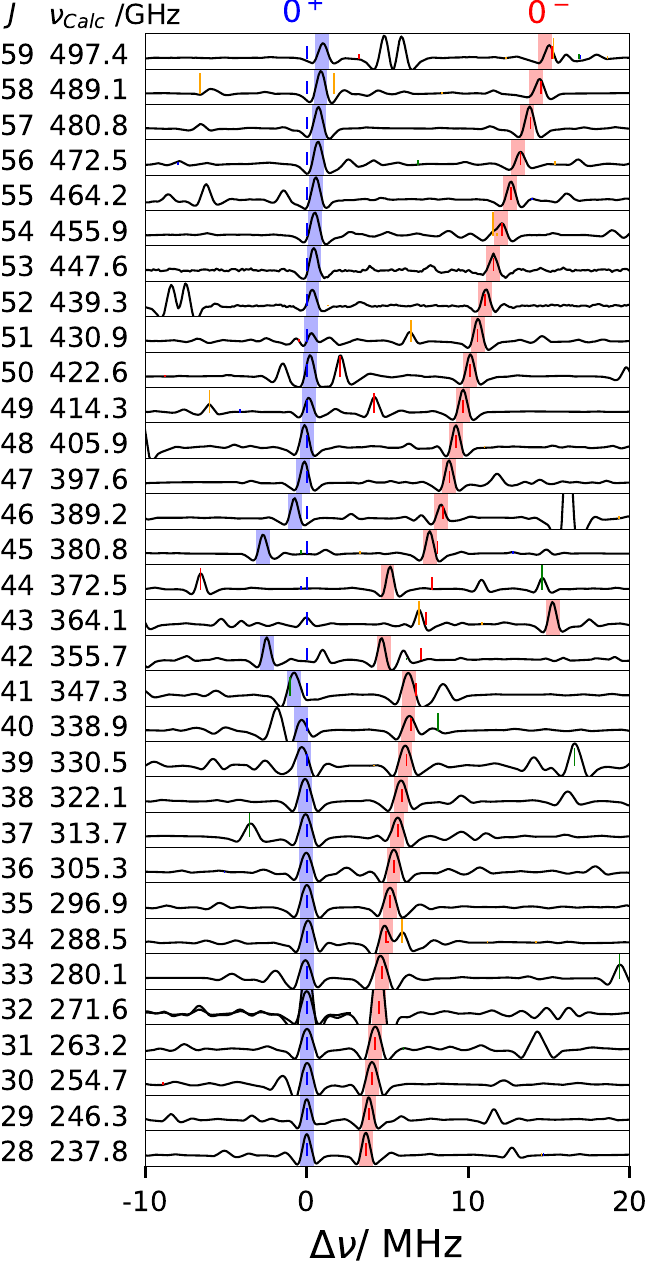}
  \end{minipage}
\begin{minipage}[l]{.226\textwidth}
  \includegraphics[width=\linewidth]{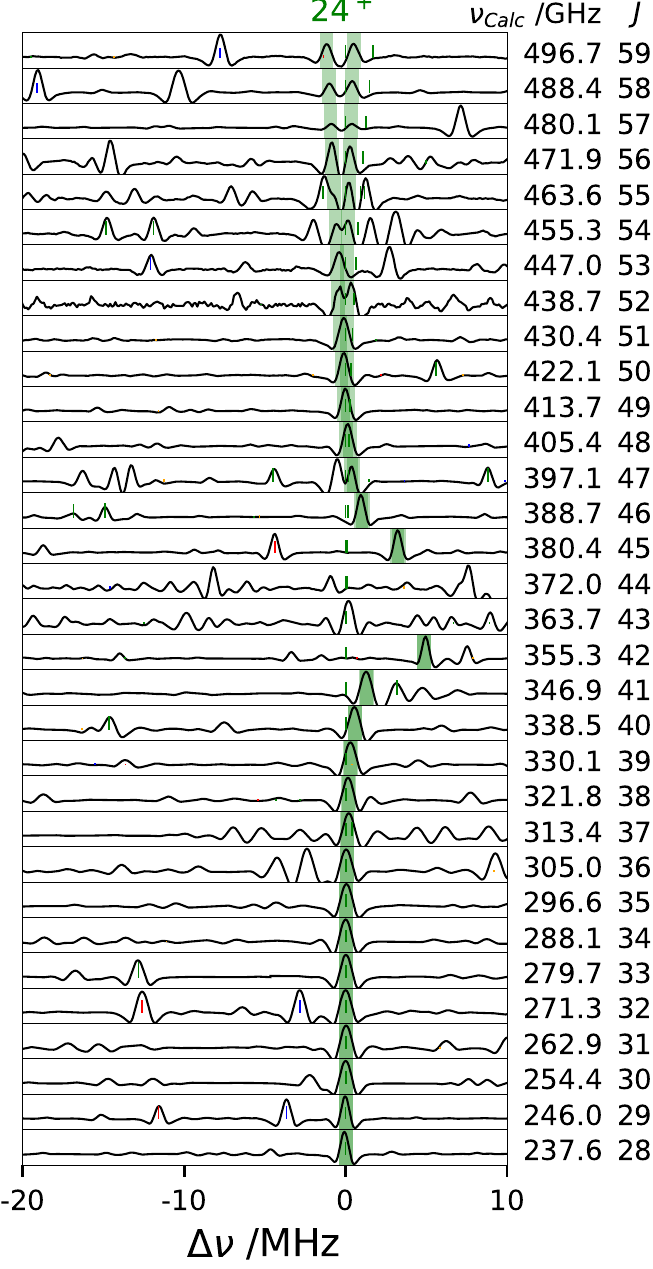}
  \end{minipage}
\caption[Loomis Wood plots of \textit{gauche}-propanal showing $^qR$ series with $K_a=12$ ($0^+$ and $0^-$) and $K_a=7$ ($24^+$).]
{
Loomis Wood plots of \textit{gauche}-propanal showing $^qR$ series transitions, $J_{K_a,K_c}\leftarrow(J-1)_{K_a,K_c-1}$, with $K_a=12$ ($0^+$ and $0^-$) on the left and $K_a=7$ ($24^+$) on the right hand side.
The predicted frequencies are derived from the respective single vibrational state models (shown also as stick spectra).
The trends in deviations from predicted to observed frequencies (final assignments are marked by colored boxes) have opposite sign for the two vibrational states. The eight strongest perturbed transitions (with $J=43$ or $J=44$), four allowed plus four nominally forbidden transitions with deviations up to 75\,MHz, are off-scale here but depicted in Fig.~\ref{FigA:4_Ka5_transitions}.
The systematic deviations are indicating interactions with $\Delta K_a=5$ between $\upsilon=0$ and $\upsilon_{24}=1$ and the incorrect determination of the fundamental frequency $\nu_{24}=113.1(2)$\,cm$^{-1}$~in the literature~\cite{Propanal_8_FIR, Propanal_7_PotentialFunction}, see text.
}
\label{Fig:7_LW_gauche_both}
\end{figure}

In the single vibrational state analysis of $\upsilon_{24}=1$, see Sec~\ref{SubSec:v24}, conspicuous deviations are observed for $K_a=7$ and $K_a=10$ which appear at the same $J$'s as deviations in $K_a=12$ and $K_a=14$ of $\upsilon=0$, respectively. 
Furthermore, the deviations seem to have opposite sign for the two vibrational states.
This is examplified by
Loomis Wood plots for $^qR$ series of $K_a=12$ ($\upsilon=0$) and of $K_a=7$ ($\upsilon_{24}=1$) in Fig.~\ref{Fig:7_LW_gauche_both}.

\begin{figure}[t]
\centering
\includegraphics[width=0.8\linewidth]{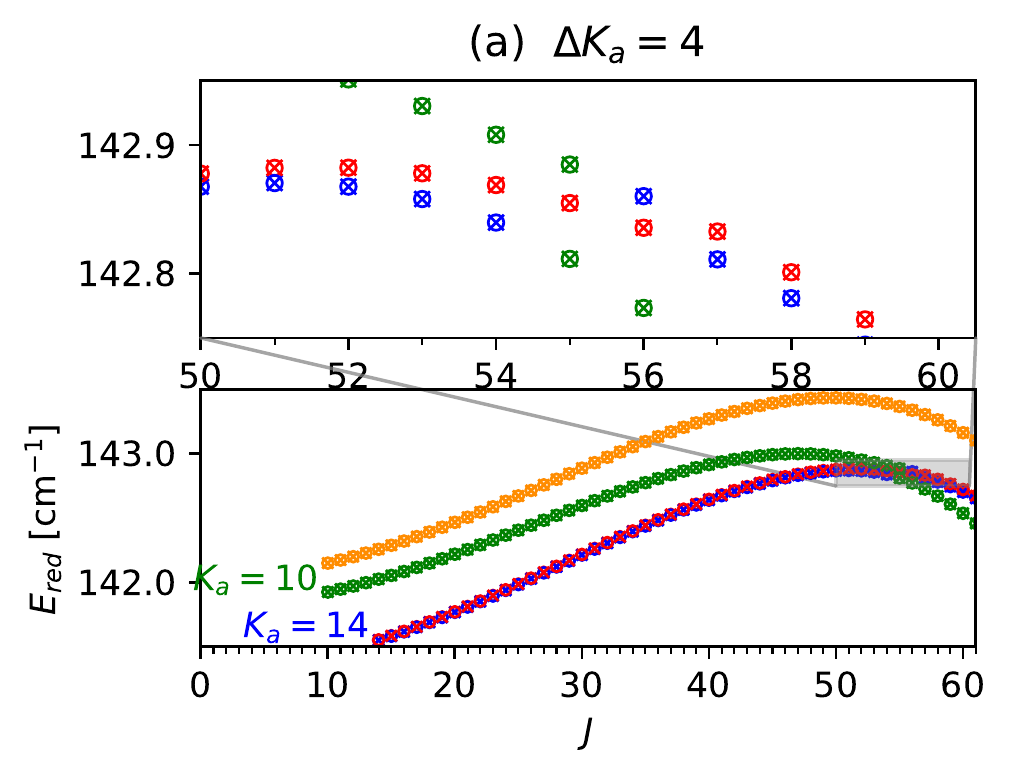}
\includegraphics[width=0.8\linewidth]{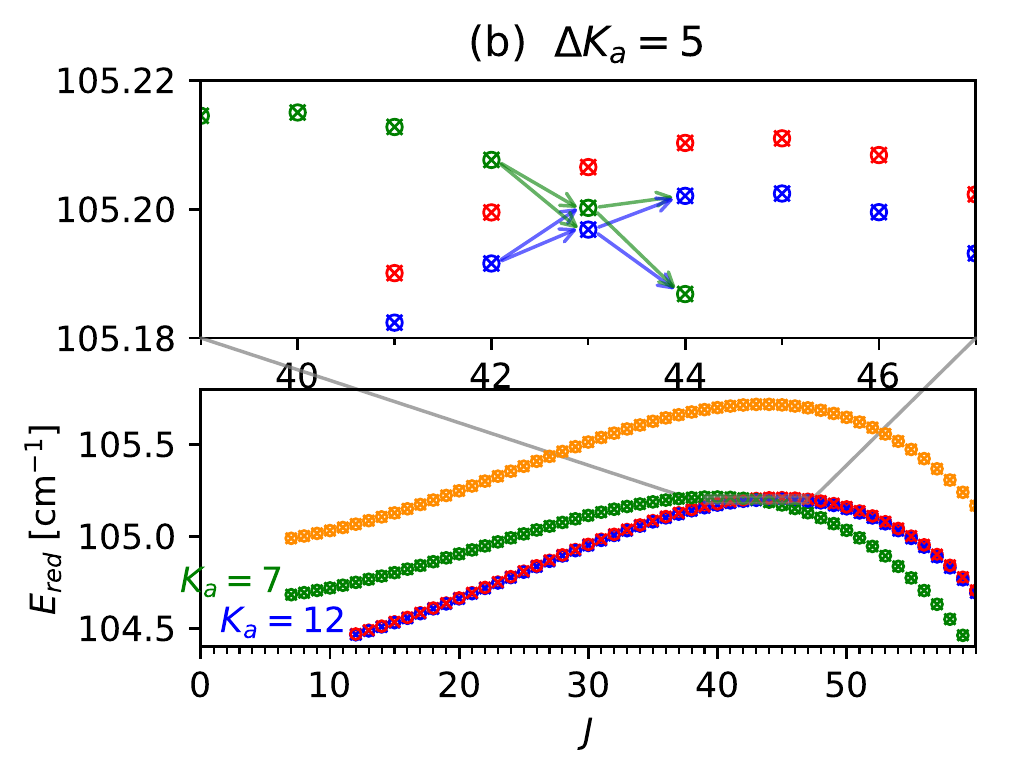}
\caption{
Enlarged portions of the reduced energy diagram of \textit{gauche}-propanal, cf. Fig.~\ref{Fig:5_Re-overview}. Highlighted are the (a) $\Delta K_a=4$ and (b) $\Delta K_a=5$ interaction areas. The assignment of the four allowed and four nominally forbidden transitions marked by the eight arrows in (b) are plotted additionally in Fig.~\ref{FigA:4_Ka5_transitions} of the supplementary material. The importance of these transitions is explained in the text.
}
\label{Fig:8_Red-Egy-interactions}
\end{figure}

In fact, Fermi resonance and Coriolis interaction between the two vibrational states are the cause of the perturbed transitions.
However, the fundamental frequency needs to be drastically reduced from $\nu_{24}=113.1$\,cm$^{-1}$ \cite{Propanal_8_FIR,Propanal_7_PotentialFunction} to be able to model the perturbed transitions to experimental accuracy.
An initial vibrational energy of $\upsilon_{24}=1$ was obtained by a careful, manual adjustment. A reduction by more than 44\,cm$^{-1}$ to about 68.8\,cm$^{-1}$ brought observed interacting energy levels close in energy, i.e. energy levels of $K_a=12$ of $\upsilon=0$ with $K_a=7$ of $\upsilon_{24}=1$ as well as $K_a=14$ of $\upsilon=0$ with $K_a=10$ of $\upsilon_{24}=1$.
The new energy difference compares very favorably with results from quantum-chemical calculations performed here at the CCSD(T)/ANO0 level of theory which yield a fundamental wavenumber of $\nu_{24}=74$\,cm$^{-1}$.
On the other hand, simple microwave relative intensity measurements \cite{Propanal_7_PotentialFunction} agreed with a higher frequency of $\nu_{24}=113.1$\,cm$^{-1}$ \cite{Propanal_8_FIR,Propanal_7_PotentialFunction}, for that reason new millimeter-wave relative intensity measurements of 48 arbitrary chosen rotational lines are performed. In general, intensity comparisons of absorption spectrometers should be treated with caution, nevertheless, they reveal a ratio of $I_{\upsilon_{24}=1}/I_{\upsilon=0}=0.76(9)$ (Table~\ref{TabA:MMW_Rel_Intensity_v24} in the supplementary material), that translates via the Boltzmann factor to $57^{+27}_{-24}$\,cm$^{-1}$, and support the newly derived value. 
The resulting $\Delta K_a=5$ and $\Delta K_a=4$ interaction areas are marked in turquoise in Fig.~\ref{Fig:5_Re-overview}. 
Reduced energy diagrams of the close up of the two interaction areas are shown in Fig.~\ref{Fig:8_Red-Egy-interactions} for clarification.
Strong perturbations occur since energy levels of $24^+$ are crossing $0^+$ and $0^-$ ones.
At this point, all parameters accounting for possible interaction between the three aforementioned tunneling states are tested and the ones ameliorating the root-mean square ($rms$) of the fit the most are incorporated into the parameter set.
The five lowest order parameters are
examined to find the best fit for each Coriolis interaction type, cf. order $n$ in Eq.~\ref{Eq:Coriolis_Terms_specific}.
For example, Fermi resonance between $0^+$ and $24^+$ are satisfactorily reproducing the observed $\Delta K_a=5$ interaction.

Four nominally forbidden $^qR$ series transitions with $J=43$ or $J=44$ ($0^+\leftrightarrow24^+$) in addition to four allowed ones ($0^+\leftrightarrow0^+$ and $24^+\leftrightarrow24^+$) are observable on account of mixed wave functions, due to the Fermi resonance. These are depicted by the eight arrows in Fig.~\ref{Fig:8_Red-Egy-interactions}b and eventually fit to experimental accuracy (cf. also Fig.~\ref{FigA:4_Ka5_transitions} in the supplementary material).
In particular the four nominally forbidden transitions between the two vibrational states give direct information on the fundamental frequency $\nu_{24}$.
The simulated interaction of energy levels can be seen by their mixing coefficients, which are visualized in Fig.~\ref{FigA:1_mixing} in the supplementary material.

The spectroscopic parameters of the $\upsilon=0$ and $\upsilon_{24}=1$ system of \textit{gauche}-propanal can be found in Table~\ref{Tab:Spectroscopic_parameters}.
In addition, the line list (\textit{*.lin}), containing all assigned transitions with their respective quantum numbers, together with the parameters setting up the Hamiltonian in Eq.~\eqref{Eq:Hamiltonian_Full} (\textit{*.par}) and the resulting fit (\textit{*.fit}) are provided in the supplementary material.
Finally, $^qR$ series up to $J_{max}=60$ and $K_a = 15$ of $\upsilon=0$ and $K_a = 11$ of $\upsilon_{24}=1$ are fit to experimental accuracy.
The total quantum number coverage of transitions and their deviations to predictions, based on parameters in Table~\ref{Tab:Spectroscopic_parameters}, can be seen in Fig.~\ref{FigA:2_QN} in the supplementary material.
The observed Fermi resonance and Coriolis interaction allow to accurately determine the fundamental frequency of $\nu_{24}=68.75037(30)$\,cm$^{-1}$.

\setlength{\tabcolsep}{0pt}
\begin{table*}
\centering
\caption[Spectroscopic parameters of the global fit of $\upsilon=0$ and $\upsilon_{24}=1$ of \textit{gauche}-propanal.]{Spectroscopic parameters$^a$~~(MHz) of the global fit of $\upsilon=0$ and $\upsilon_{24}=1$ of \textit{gauche}-propanal.}
\label{Tab:Spectroscopic_parameters}
\tiny
\begin{tabular}{lllrlrlrlrl}
\toprule
$\hat{H}$~~~~~~~ & Parameter & & & $0^+$ &  & $0^-$ & & $24^+$ & & $24^-$  \\
\midrule
&${\Delta E_\upsilon/2}^c$    &  &  \multicolumn{4}{c}{237.7893( 81)}  & \multicolumn{4}{c}{6708.3692(235)} \\ 
&$E_{24}$ &       &  \multicolumn{8}{c}{2061084.2(90)}   \\
~\\
&${E_{rel}}^f$ &  &  $-$237&.7893( 81) &  237&.7893( 81)   &   2054375&.8( 90) & 2067792&.6( 90)  \\
&${E_{rel}}^f$ &/~cm$^{-1}$   &  $-$0&.00793180( 27)~ & 0&.00793180( 27)~ & 68&.52660( 30)~ & 68&.97414( 30)  \\
&${\Delta E_\upsilon}^f$         & &  \multicolumn{4}{c}{475.5786(115)} & \multicolumn{4}{c}{13416.8(13)}  \\
&${\Delta E_\upsilon}^f$ &/~cm$^{-1}$  &  \multicolumn{4}{c}{0.01586359(38)} & \multicolumn{4}{c}{0.44753(42)}  \\
&${E_{24}}^f$ &/~cm$^{-1}$  &  \multicolumn{8}{c}{68.75037(30)}   \\
~\\
$\hat{H}_\upsilon$ & $A$ & & 26252&.5958( 75) & 26248&.5661( 79) & 26460&.7047(155) & 26355&.6568(204)  \\
&$B$ & & 4314&.88778( 67) & 4314&.95158( 62) & 4326&.01733( 78) & 4329&.30966( 59)  \\
&$C$ & & 4147&.91471( 61) & 4148&.12679( 59) & 4159&.09496(106) & 4164&.83173( 96)  \\

&$-D_{K}$ &          & $-$2&.06977( 62)   & $-$2&.05824( 63)   & $-$2&.49135(150)  & $-$2&.0896( 39)   \\
&$-D_{JK}$&          &  0&.1798841( 50) &  0&.1789892( 47) &  0&.218507( 51) &  0&.188766( 78)  \\
&$-D_{J}$ &$\times10^3$  & $-$6&.580854(308)  & $-$6&.56201( 32)   & $-$7&.90169( 37)  & $-$7&.26255( 53)  \\
&$d_1   $ &$\times10^3$  &  1&.08499( 57)   &  1&.07526( 57)   &  1&.31865( 45)  &  1&.14660( 37)   \\
&$d_2   $ &$\times10^3$  & $-$0&.044125( 68)  & $-$0&.039849( 38)  &  0&.038543(150) &  0&.063808( 93)  \\

&$H_{K} $ &$ \times10^6$ & $-$& & $-$& & $-$& & $-$&  \\
&$H_{KJ}$ &$ \times10^6$ &  8&.709( 32)    & 11&.200( 34)    & $-$83&.926(268)    &  5&.338(290)  \\
&$H_{JK}$ &$ \times10^6$ & $-$3&.2406( 35)   & $-$3&.3057( 34)   &   1&.1632( 76)   & $-$2&.2458( 99)  \\
&$H_{J} $ &$ \times10^6$ &  0&.127377(111) &  0&.128224(113) &   0&.092240(215) &  0&.138476(219)  \\
&$h_1   $ &$\times10^9$  & $-$51&.703(158) & $-$52&.556(156) & $-$9&.608( 76) & $-$32&.140( 55)  \\
&$h_2   $ &$\times10^9$  & $-$0&.8131( 85) & $-$& & $-$23&.001( 63)$^{d}$ & $-$23&.001( 63)$^{d}$  \\
&$h_3   $ &$\times10^9$  & $-$& & $-$0&.19834(244) & 2&.2481( 57) & 0&.5638(248)  \\

&$L_{K}  $ &$  \times10^9$ & $-$& & $-$& & $-$& & $-$&  \\
&$L_{KKJ}$ &$  \times10^9$ & $-$2&.265( 87)   &   $-$5&.299( 88)   & 94&.94(142)    &  $-$&  \\
&$L_{JK} $ &$  \times10^9$ & $-$2&.3217(182)  &   $-$2&.2200(182)  & $-$1&.522( 75)   &  2&.476( 81)  \\
&$L_{JJK}$ &$  \times10^9$ &  0&.14531( 87) &    0&.14138( 88) & $-$0&.15456(145) & $-$0&.16335(225)  \\
&$L_{J}  $ &$  \times10^{12}$ & $-$3&.3327(284) & $-$3&.2059(284)  & $-$3&.864( 60)   & $-$2&.458( 60)  \\

&$l_1   $ &$\times10^{12}$ & 2&.1378(128) & 2&.0364(128) & $-$& & $-$&  \\
&$l_2   $ &$\times10^{12}$ & $-$& & $-$& & 1&.0502(103) & 3&.3197(278)  \\
&$l_3   $ &$\times10^{12}$ & $-$& & $-$& & $-$& & $-$0&.2006( 61)  \\
&$l_4   $ &$\times10^{12}$ & $-$& & $-$& & $-$& & $-$&  \\

&$P_{JK} $ &$\times10^{12}$ & $-$0&.15475(281)$^{d}$ & $-$0&.15475(281)$^{d}$ & $-$0&.2404(105) & $-$0&.3999(132)  \\
&$P_J    $ &$\times10^{15}$ & 0&.07456(286)$^{d}$ & 0&.07456(286)$^{d}$ & 0&.1630( 62) & 0&.0813( 63)  \\
&$p_2    $ &$\times10^{15}$ & $-$& & $-$& & $-$& & $-$0&.2204( 40)  \\
~\\
$\hat{H}^a_{\upsilon^\pm}$ &$F_{bc}$     &            & \multicolumn{4}{c}{23.96486(229)} & \multicolumn{4}{c}{32.22732( 54)}  \\
&$F_{bc}^{K}$ &            & \multicolumn{4}{c}{$-$0.10958( 81)} & \multicolumn{4}{c}{0.399567(297)}  \\
&$F_{bc}^{J}$ &$\times10^3$ & \multicolumn{4}{c}{$-$3.3607( 40)}  & \multicolumn{4}{c}{$-$2.96460(216)}  \\
&$F_{bc}^{KK}$&$\times10^3$ & \multicolumn{4}{c}{0.24036( 86)}  & \multicolumn{4}{c}{$-$} \\
&$F_{bc}^{JK}$&$\times10^3$ & \multicolumn{4}{c}{$-$0.013551(106)}& \multicolumn{4}{c}{$-$0.023756( 97)}  \\
&$F_{bc}^{JJ}$&$\times10^6$ & \multicolumn{4}{c}{0.14945( 63)}  & \multicolumn{4}{c}{0.07229( 40)}  \\
&$F_{bc}^{JJK}$&$\times10^9$& \multicolumn{4}{c}{$-$}             & \multicolumn{4}{c}{0.5455(163)}  \\
&$F_{bc}^{KKK}$&$\times10^6$& \multicolumn{4}{c}{0.1473( 39)} & \multicolumn{4}{c}{$-$}  \\
&$F_{2bc}$   &$\times10^3$ & \multicolumn{4}{c}{$-$0.34902( 47)} & \multicolumn{4}{c}{$-$}  \\
~\\
$\hat{H}^b_{\upsilon^\pm}$&$F_{ca}^{}$  &            & \multicolumn{4}{c}{$-$} & \multicolumn{4}{c}{151.909( 61)} \\
&$F_{ca}^{K}$ &            & \multicolumn{4}{c}{$-$} & \multicolumn{4}{c}{$-$0.1580( 32)} \\
&$F_{ca}^{J}$ &$\times10^3$ & \multicolumn{4}{c}{$-$} & \multicolumn{4}{c}{$-$4.489( 45)}  \\
&$F_{ca}^{KK}$&$\times10^3$ & \multicolumn{4}{c}{$-$} & \multicolumn{4}{c}{$-$0.4677( 61)} \\
&$F_{ca}^{JJ}$&$\times10^6$ & \multicolumn{4}{c}{$-$} & \multicolumn{4}{c}{0.2043( 78)}  \\
~\\
$\hat{H}^c_{\upsilon^\pm}$ &${F_{ab}/2}^c$ &            & \multicolumn{4}{c}{5.17( 38)}& \multicolumn{4}{c}{4.3454(139)} \\
&${F_{ab}^{K}/2}^c$ &            & \multicolumn{4}{c}{0.066$^e$}    & \multicolumn{4}{c}{$-$} \\
&${F_{ab}^{J}/2}^c$ &$\times10^3$ & \multicolumn{4}{c}{$-$}        & \multicolumn{4}{c}{1.1419( 57)} \\
&${F_{ab}^{JJ}/2}^c$&$\times10^6$ & \multicolumn{4}{c}{$-$}        & \multicolumn{4}{c}{0.10361(207)} \\
~\\
$\hat{H}^C$&${F_{bc}}^g$   &            & \multicolumn{8}{c}{7.2891(115)}     \\
&${G_{2b}}^g$  &            & \multicolumn{8}{c}{0.020888(188)}   \\
&${G_{2b}^J}^g$  &$\times10^6$ & \multicolumn{8}{c}{2.906( 57)}      \\
$\hat{H}^F_-$&\multicolumn{2}{l}{${F_2^-}^h$}  & \multicolumn{8}{c}{$-$3.1298( 64)}    \\
$\hat{H}^F_+$&\multicolumn{2}{l}{${F_2^+}^h$}   & \multicolumn{8}{c}{$-$2.6569(122)}    \\
\hline
&\multicolumn{2}{l}{Global fit} \\
&\multicolumn{2}{l}{Number of transitions$^b$} & 1414 && 1419 && 985 && 992\\
&\multicolumn{2}{l}{Total number of transitions} & & & 4810\\
&\multicolumn{2}{l}{Total number of lines}       & & &3426\\

&$rms$ &/ kHz &  & & 55&.2\,kHz \\
&$wrms$ &/ unitless & & & 1&.17 \\
\bottomrule
\bottomrule

\end{tabular}
\begin{flushleft}
    \vspace{0.5em}
    {$^a$ S reduction is used with SPFIT in the $I^{r}$-representation. Values of $F_{bc}$ etc., given either between column $0^+$ and $0^-$ or $24^+$ and $24^-$, are associated with operators between tunneling states. Values of $F_{bc}$ etc., given between column $0^-$ and $24^+$, are associated with operators between the vibrational states. The column $\hat{H}$ denotes to which Hamiltonian of Eq.~\ref{Eq:Hamiltonian_Full} the parameters from one block belong to.\\
$^b$ Transitions between different tunneling states are counted to the respective lower level of a transition.\\
$^c $ Parameter $X$ is fit as linear combination $X=(X^{\upsilon'}-X^{\upsilon''})/2$.\\
$^{d}$ Parameter $X$ is fit as linear combination $X=(X^{\upsilon''}+X^{\upsilon'})/2$.\\
$^e$ Parameter is manually fixed.\\
$^f$ Value not fit. Only given for clarification and calculated from the fit $\Delta E_\upsilon/2$ and $ E_{24}$ parameters.\\ 
$^g$ Parameter acts between $0^+$ and $24^-$ as well as between $0^-$ and $24^+$.\\
$^h$ Parameter acts between $0^+$ and $24^+$ (+) or between $0^-$ and $24^-$ ($-$).
}
\end{flushleft}
\end{table*}

\section{Discussion and Prospects}            
\label{Sec:Discussion}                        

The main result of the combined analysis of $\upsilon=0$ and $\upsilon_{24}=1$ is the re-determination of the fundamental frequency of the aldehyde torsion to $\nu_{24}=68.75037(30)$\,cm$^{-1}$.
This result is further verified by high-level quantum-chemical calculations and millimeter-wave relative intensity measurements.
In addition, the redetermined fundamental frequency $\nu_{24}$ of \textit{gauche}-propanal is comparable to 70\,cm$^{-1}$ of \textit{gauche}-nitrosoethane \cite{Propanal_7_PotentialFunction}, its isoelectronic molecule (CH$_3$CH$_2$NO).
Similar fundamental frequencies were already observed for their \textit{syn} conformers, with \textit{syn}-propanal of $\nu_{24}=135.1$\,cm$^{-1}$ \cite{Propanal_8_FIR} and \textit{syn}-nitrosoethane of 130\,cm$^{-1}$ \cite{Propanal_7_PotentialFunction}. These qualitative agreements of fundamental frequencies may additionally support the re-determination.
The energy splitting of the respective tunneling states of a vibrational state is increasing from about 16$\times10^{-3}$\,cm$^{-1}$ ($\upsilon=0$) to 448$\times10^{-3}$\,cm$^{-1}$ ($\upsilon_{24}=1$). A larger splitting is commonly observed for energetically higher excited vibrational states below the barrier of a double potential well.

Assigning ro-vibrational spectra in the far-infrared (FIR) region in the future would verify the energy differences by a direct measurement of the new fundamental frequency $\nu_{24}$.
Another approach for further verifying $\nu_{24}$ are very accurate relative intensity measurements with the help of emission spectroscopy, which allows to detect absolute line intensities \cite{wehres2017_emission,wehres2018_emission}.
A revisit of the fundamental frequency $\nu_{24}$ of \textit{syn}-propanal using high-resolution FIR spectroscopy is recommended considering the re-determination for \textit{gauche}-propanal presented here. The fundamental frequencies of the aldehyde torsion of \textit{syn}- and \textit{gauche}-propanal, in the best case in combination with additional frequencies for overtone and hot bands, allow to accurately model the interconversion potential, or the $-$CHO-group rotation, of propanal in the future. 

The present analysis profited immensely from DM-DR spectroscopy in the W-band region \cite{Zingsheim2021_DMDR} for three reasons: i) linkages allowed for straightforward assignments in the W-band region without the need of \textit{ab-initio} calculations, ii) strongly perturbed transitions (with $K_a=1$) could be assigned unambiguously (cf. Fig.~\ref{Fig:3_Fortrat-DR-links}) and iii) weak $c$-type transitions connecting the two tunneling states were assigned easily (cf. Fig.~\ref{Fig:6_c-types}). In particular the latter reason may also allow to improve the analyses of rotational spectra of molecules with two stable degenerate \textit{gauche} or \textit{synclinical} conformers in the future. 

Increasing the frequency range above 500\,GHz could help to even better describe the $\Delta K_a=4$ interaction, as in this case unperturbed transitions with $J>60$ may be beneficial for the modeling of the interaction.
Furthermore, extending the DM-DR measurements beyond the W-band would allow for assignments of perturbed transitions in a straightforward fashion. Ideally, the eight transitions marked by the arrows in Fig.~\ref{Fig:8_Red-Egy-interactions}b and plotted in Fig.~\ref{FigA:4_Ka5_transitions} in the supplementary material could be additionally confirmed by DR spectroscopy, as these transitions contain the most information on the $\Delta K_a=5$ interaction and could secure the assignments and confirm the spectroscopic model even further.
Increasing the covered quantum number range, cf. Fig.~\ref{FigA:2_QN} in the supplemental material, is only advisable if the second excited aldehyde torsion is included in a global analysis to be able to properly treat perturbed transitions originating from interactions between $\upsilon=0$ or $\upsilon_{24}=1$ with $\upsilon_{24}=2$. Furthermore, by including $\upsilon_{24}=2$ in the rotational analysis, an assignment of the first overtone band of the aldehyde torsion $\upsilon_{24}=2\leftarrow0$ may be straightforward and additionally confirm the newly determined fundamental frequency $\nu_{24}$ indirectly. 

In conclusion, the sophisticated rotational analysis of $\upsilon=0$ and $\upsilon_{24}=1$ of \textit{gauche}-propanal allows astronomers to search for its rotational fingerprints in space and may guide the analysis of FIR spectra to even better understand the LAMs of propanal.

\section*{Acknowledgement}
This work has been supported via Collaborative Research Centre 956, sub-project B3, funded by the Deutsche Forschungsgemeinschaft  (DFG; project ID 184018867) and DFG SCHL 341/15-1 
(``Cologne Center for Terahertz Spectroscopy"). 

\bibliographystyle{elsarticle-num-names}
\bibliography{gauche-Propanal}

\begin{thebibliography}{39}
\expandafter\ifx\csname natexlab\endcsname\relax\def\natexlab#1{#1}\fi
\providecommand{\url}[1]{\texttt{#1}}
\providecommand{\href}[2]{#2}
\providecommand{\path}[1]{#1}
\providecommand{\DOIprefix}{doi:}
\providecommand{\ArXivprefix}{arXiv:}
\providecommand{\URLprefix}{URL: }
\providecommand{\Pubmedprefix}{pmid:}
\providecommand{\doi}[1]{\href{http://dx.doi.org/#1}{\path{#1}}}
\providecommand{\Pubmed}[1]{\href{pmid:#1}{\path{#1}}}
\providecommand{\bibinfo}[2]{#2}
\ifx\xfnm\relax \def\xfnm[#1]{\unskip,\space#1}\fi
\bibitem[{Butcher and Wilson(1964)}]{Propanal_1_Butcher_Wilson}
\bibinfo{author}{S.~S. Butcher}, \bibinfo{author}{E.~B. Wilson},
\newblock \bibinfo{title}{Microwave spectrum of propionaldehyde},
\newblock \bibinfo{journal}{J. Chem. Phys.} \bibinfo{volume}{40}
  (\bibinfo{year}{1964}) \bibinfo{pages}{1671--1678}.
  \DOIprefix\doi{10.1063/1.1725377}.
\bibitem[{Pickett and Scroggin(1974)}]{Propanal_2_Pickett_Scroggin}
\bibinfo{author}{H.~M. Pickett}, \bibinfo{author}{D.~G. Scroggin},
\newblock \bibinfo{title}{{Microwave spectrum and internal rotation potential
  of propanal}},
\newblock \bibinfo{journal}{J. Chem. Phys.} \bibinfo{volume}{61}
  (\bibinfo{year}{1974}) \bibinfo{pages}{3954--3958}.
  \DOIprefix\doi{10.1063/1.1681688}.
\bibitem[{Hardy et~al.(1982)Hardy, Cox, Fliege, and Dreizler}]{Propanal_3_JAH}
\bibinfo{author}{J.~A. Hardy}, \bibinfo{author}{A.~P. Cox},
  \bibinfo{author}{E.~Fliege}, \bibinfo{author}{H.~Dreizler},
\newblock \bibinfo{title}{{Determination of a High Potential Barrier Hindering
  Internal Rotation from the Ground State Spectrum The Methylbarrier of
  cis-Propanal}},
\newblock \bibinfo{journal}{Z. Naturforsch. A} \bibinfo{volume}{37}
  (\bibinfo{year}{1982}) \bibinfo{pages}{1035 -- 1037}.
  \DOIprefix\doi{10.1515/zna-1982-0910}.
\bibitem[{Randell et~al.(1987)Randell, Cox, and
  Dreizler}]{Propanal_4_Randell_Gauche}
\bibinfo{author}{J.~Randell}, \bibinfo{author}{A.~P. Cox},
  \bibinfo{author}{H.~Dreizler},
\newblock \bibinfo{title}{{Gauche Propanal: Microwave Spectrum and Methyl
  Barrier}},
\newblock \bibinfo{journal}{Z. Naturforsch. A} \bibinfo{volume}{42}
  (\bibinfo{year}{1987}) \bibinfo{pages}{957 -- 962}.
  \DOIprefix\doi{10.1515/zna-1987-0908}.
\bibitem[{Demaison et~al.(1987)Demaison, Maes, {Van Eijck}, Wlodarczak, and
  Lasne}]{Propanal_5_Demaison}
\bibinfo{author}{J.~Demaison}, \bibinfo{author}{H.~Maes},
  \bibinfo{author}{B.~P. {Van Eijck}}, \bibinfo{author}{G.~Wlodarczak},
  \bibinfo{author}{M.~C. Lasne},
\newblock \bibinfo{title}{{Determination of the moment of inertia of methyl
  groups: Analysis of the millimeterwave spectra of \textit{cis}-propanal and
  methylthioethyne}},
\newblock \bibinfo{journal}{J. Mol. Spectrosc.} \bibinfo{volume}{125}
  (\bibinfo{year}{1987}) \bibinfo{pages}{214 -- 224}.
  \DOIprefix\doi{10.1016/0022-2852(87)90208-6}.
\bibitem[{Randell et~al.(1988{\natexlab{a}})Randell, Cox, ii~Hillig, Imachi,
  LaBarge, and Kuczkowski}]{Propanal_6_Structure}
\bibinfo{author}{J.~Randell}, \bibinfo{author}{A.~P. Cox},
  \bibinfo{author}{K.~W. ii~Hillig}, \bibinfo{author}{M.~Imachi},
  \bibinfo{author}{M.~S. LaBarge}, \bibinfo{author}{R.~L. Kuczkowski},
\newblock \bibinfo{title}{{Cis and Gauche Propanal: Microwave Spectra and
  Molecular Structures}},
\newblock \bibinfo{journal}{Z. Naturforsch. A} \bibinfo{volume}{43}
  (\bibinfo{year}{1988}{\natexlab{a}}) \bibinfo{pages}{271 -- 276}.
  \DOIprefix\doi{10.1515/zna-1988-0314}.
\bibitem[{Randell et~al.(1988{\natexlab{b}})Randell, Hardy, and
  Cox}]{Propanal_7_PotentialFunction}
\bibinfo{author}{J.~Randell}, \bibinfo{author}{J.~A. Hardy},
  \bibinfo{author}{A.~P. Cox},
\newblock \bibinfo{title}{{The microwave spectrum and potential function of
  propanal}},
\newblock \bibinfo{journal}{J. Chem. Soc.{,} Faraday Trans. 2}
  \bibinfo{volume}{84} (\bibinfo{year}{1988}{\natexlab{b}})
  \bibinfo{pages}{1199--1212}. \DOIprefix\doi{10.1039/F29888401199}.
\bibitem[{Durig et~al.(1980)Durig, Compton, and McArver}]{Propanal_8_FIR}
\bibinfo{author}{J.~R. Durig}, \bibinfo{author}{D.~A.~C. Compton},
  \bibinfo{author}{A.~Q. McArver},
\newblock \bibinfo{title}{{Low frequency vibrational spectra, methyl torsional
  potential functions, and internal rotational potential of propanal}},
\newblock \bibinfo{journal}{J. Chem. Phys.} \bibinfo{volume}{73}
  (\bibinfo{year}{1980}) \bibinfo{pages}{719--724}.
  \DOIprefix\doi{10.1063/1.440173}.
\bibitem[{Guirgis et~al.(1998)Guirgis, Drew, Gounev, and Durig}]{Propanal_9_IR}
\bibinfo{author}{G.~A. Guirgis}, \bibinfo{author}{B.~R. Drew},
  \bibinfo{author}{T.~K. Gounev}, \bibinfo{author}{J.~R. Durig},
\newblock \bibinfo{title}{{Conformational stability and vibrational assignment
  of propanal}},
\newblock \bibinfo{journal}{Spectrochim. Acta A Mol. Biomol. Spectrosc.}
  \bibinfo{volume}{54} (\bibinfo{year}{1998}) \bibinfo{pages}{123 -- 143}.
  \DOIprefix\doi{10.1016/S1386-1425(97)00200-X}.
\bibitem[{Zingsheim et~al.(2017)Zingsheim, M\"uller, Lewen, J{\o}rgensen, and
  Schlemmer}]{ZINGSHEIM2017_Propanal}
\bibinfo{author}{O.~Zingsheim}, \bibinfo{author}{H.~S.~P. M\"uller},
  \bibinfo{author}{F.~Lewen}, \bibinfo{author}{J.~K. J{\o}rgensen},
  \bibinfo{author}{S.~Schlemmer},
\newblock \bibinfo{title}{Millimeter and submillimeter wave spectroscopy of
  propanal},
\newblock \bibinfo{journal}{J. Mol. Spectrosc.} \bibinfo{volume}{342}
  (\bibinfo{year}{2017}) \bibinfo{pages}{125 -- 131}.
  \DOIprefix\doi{10.1016/j.jms.2017.07.008}.
\bibitem[{Hollis et~al.(2004)Hollis, Jewell, Lovas, Remijan, and
  M{\o}llendal}]{First_Detection_Propanal}
\bibinfo{author}{J.~M. Hollis}, \bibinfo{author}{P.~R. Jewell},
  \bibinfo{author}{F.~J. Lovas}, \bibinfo{author}{A.~Remijan},
  \bibinfo{author}{H.~M{\o}llendal},
\newblock \bibinfo{title}{{Green Bank Telescope Detection of New Interstellar
  Aldehydes: Propenal and Propanal}},
\newblock \bibinfo{journal}{ApJ} \bibinfo{volume}{610} (\bibinfo{year}{2004})
  \bibinfo{pages}{L21--L24}. \DOIprefix\doi{10.1086/423200}.
\bibitem[{Requena-Torres et~al.(2008)Requena-Torres, Martin-Pintado, Martin,
  and Morris}]{Requena_Torres_2008}
\bibinfo{author}{M.~A. Requena-Torres}, \bibinfo{author}{J.~Martin-Pintado},
  \bibinfo{author}{S.~Martin}, \bibinfo{author}{M.~R. Morris},
\newblock \bibinfo{title}{The galactic center: The largest oxygen-bearing
  organic molecule repository},
\newblock \bibinfo{journal}{ApJ} \bibinfo{volume}{672} (\bibinfo{year}{2008})
  \bibinfo{pages}{352--360}. \DOIprefix\doi{10.1086/523627}.
\bibitem[{Lykke et~al.(2017)Lykke, Coutens, J\o{}rgensen, van~der Wiel, Garrod,
  M\"uller, Bjerkeli, Bourke, Calcutt, Drozdovskaya, Favre, Fayolle, Jacobsen,
  \"Oberg, Persson, van Dishoeck, and Wampfler}]{Lykke_space}
\bibinfo{author}{J.~M. Lykke}, \bibinfo{author}{A.~Coutens},
  \bibinfo{author}{J.~K. J\o{}rgensen}, \bibinfo{author}{M.~H.~D. van~der
  Wiel}, \bibinfo{author}{R.~T. Garrod}, \bibinfo{author}{H.~S.~P. M\"uller},
  \bibinfo{author}{P.~Bjerkeli}, \bibinfo{author}{T.~L. Bourke},
  \bibinfo{author}{H.~Calcutt}, \bibinfo{author}{M.~N. Drozdovskaya},
  \bibinfo{author}{C.~Favre}, \bibinfo{author}{E.~C. Fayolle},
  \bibinfo{author}{S.~K. Jacobsen}, \bibinfo{author}{K.~I. \"Oberg},
  \bibinfo{author}{M.~V. Persson}, \bibinfo{author}{E.~F. van Dishoeck},
  \bibinfo{author}{S.~F. Wampfler},
\newblock \bibinfo{title}{{The ALMA-PILS survey: First detections of ethylene
  oxide, acetone and propanal toward the low-mass protostar IRAS 16293-2422}},
\newblock \bibinfo{journal}{Astron. Astrophys.} \bibinfo{volume}{597}
  (\bibinfo{year}{2017}) \bibinfo{pages}{A53}. \URLprefix
  \url{https://doi.org/10.1051/0004-6361/201629180}.
  \DOIprefix\doi{10.1051/0004-6361/201629180}.
\bibitem[{Zingsheim et~al.(2021)Zingsheim, Bonah, Lewen, Thorwirth, Müller,
  and Schlemmer}]{Zingsheim2021_DMDR}
\bibinfo{author}{O.~Zingsheim}, \bibinfo{author}{L.~Bonah},
  \bibinfo{author}{F.~Lewen}, \bibinfo{author}{S.~Thorwirth},
  \bibinfo{author}{H.~S. Müller}, \bibinfo{author}{S.~Schlemmer},
\newblock \bibinfo{title}{Millimeter-millimeter-wave double-modulation
  double-resonance spectroscopy},
\newblock \bibinfo{journal}{J. Mol. Spectrosc.} \bibinfo{volume}{381}
  (\bibinfo{year}{2021}) \bibinfo{pages}{111519}.
  \DOIprefix\doi{10.1016/j.jms.2021.111519}.
\bibitem[{Martin-Drumel et~al.(2015)Martin-Drumel, van Wijngaarden, Zingsheim,
  Lewen, Harding, Schlemmer, and Thorwirth}]{Drumel2015_OSSO}
\bibinfo{author}{M.-A. Martin-Drumel}, \bibinfo{author}{J.~van Wijngaarden},
  \bibinfo{author}{O.~Zingsheim}, \bibinfo{author}{F.~Lewen},
  \bibinfo{author}{M.~E. Harding}, \bibinfo{author}{S.~Schlemmer},
  \bibinfo{author}{S.~Thorwirth},
\newblock \bibinfo{title}{{Millimeter- and submillimeter-wave spectroscopy of
  disulfur dioxide, OSSO}},
\newblock \bibinfo{journal}{J. Mol. Spectrosc.} \bibinfo{volume}{307}
  (\bibinfo{year}{2015}) \bibinfo{pages}{33 -- 39}.
  \DOIprefix\doi{10.1016/j.jms.2014.11.007}.
\bibitem[{Ordu et~al.(2019)Ordu, Zingsheim, Belloche, Lewen, Garrod, Menten,
  Schlemmer, and M\"uller}]{Ordu2019_Acetone}
\bibinfo{author}{M.~H. Ordu}, \bibinfo{author}{O.~Zingsheim},
  \bibinfo{author}{A.~Belloche}, \bibinfo{author}{F.~Lewen},
  \bibinfo{author}{R.~T. Garrod}, \bibinfo{author}{K.~M. Menten},
  \bibinfo{author}{S.~Schlemmer}, \bibinfo{author}{H.~S.~P. M\"uller},
\newblock \bibinfo{title}{{Laboratory rotational spectroscopy of isotopic
  acetone, CH$_3^{13}$C(O)CH$_3$ and $^{13}$CH$_3$C(O)CH$_3$, and astronomical
  search in Sagittarius B2(N2)}},
\newblock \bibinfo{journal}{Astron. Astrophys.} \bibinfo{volume}{629}
  (\bibinfo{year}{2019}) \bibinfo{pages}{A72}.
  \DOIprefix\doi{10.1051/0004-6361/201935887}.
\bibitem[{Autler and Townes(1955)}]{Autler_Tones_1955}
\bibinfo{author}{S.~H. Autler}, \bibinfo{author}{C.~H. Townes},
\newblock \bibinfo{title}{{Stark Effect in Rapidly Varying Fields}},
\newblock \bibinfo{journal}{Phys. Rev.} \bibinfo{volume}{100}
  (\bibinfo{year}{1955}) \bibinfo{pages}{703--722}.
  \DOIprefix\doi{10.1103/PhysRev.100.703}.
\bibitem[{Cohen-Tannoudji et~al.(2008)Cohen-Tannoudji, Dupont-Roc, and
  Grynberg}]{Tannoudji_DressedAtomApproach}
\bibinfo{author}{C.~Cohen-Tannoudji}, \bibinfo{author}{J.~Dupont-Roc},
  \bibinfo{author}{G.~Grynberg}, \bibinfo{title}{{The Dressed Atom Approach}},
  \bibinfo{publisher}{John Wiley \& Sons, Ltd}, \bibinfo{year}{2008}, pp.
  \bibinfo{pages}{407--514}. \DOIprefix\doi{10.1002/9783527617197.ch6}.
\bibitem[{Petkie et~al.(2001)Petkie, Goyette, Helminger, Pickett, and {De
  Lucia}}]{dyad_F_plus_minus}
\bibinfo{author}{D.~T. Petkie}, \bibinfo{author}{T.~M. Goyette},
  \bibinfo{author}{P.~Helminger}, \bibinfo{author}{H.~M. Pickett},
  \bibinfo{author}{F.~C. {De Lucia}},
\newblock \bibinfo{title}{{The Energy Levels of the $\nu_5/2\nu_9$ Dyad of
  HNO$_3$ from Millimeter and Submillimeter Rotational Spectroscopy}},
\newblock \bibinfo{journal}{J. Mol. Spectrosc.} \bibinfo{volume}{208}
  (\bibinfo{year}{2001}) \bibinfo{pages}{121 -- 135}.
  \DOIprefix\doi{10.1006/jmsp.2001.8367}.
\bibitem[{Kisiel et~al.(2009)Kisiel, Pszczółkowski, Drouin, Brauer, Yu, and
  Pearson}]{KISIEL200926}
\bibinfo{author}{Z.~Kisiel}, \bibinfo{author}{L.~Pszczółkowski},
  \bibinfo{author}{B.~J. Drouin}, \bibinfo{author}{C.~S. Brauer},
  \bibinfo{author}{S.~Yu}, \bibinfo{author}{J.~C. Pearson},
\newblock \bibinfo{title}{{The rotational spectrum of acrylonitrile up to 1.67
  THz}},
\newblock \bibinfo{journal}{J. Mol. Spectrosc.} \bibinfo{volume}{258}
  (\bibinfo{year}{2009}) \bibinfo{pages}{26 -- 34}.
  \DOIprefix\doi{10.1016/j.jms.2009.08.011}.
\bibitem[{Müller et~al.(2002)Müller, Cohen, and
  Christen}]{muller_cl_coriolis}
\bibinfo{author}{H.~S.~P. Müller}, \bibinfo{author}{E.~A. Cohen},
  \bibinfo{author}{D.~Christen},
\newblock \bibinfo{title}{{The Rotational Spectrum of ClClO$_2$ in Its
  $\upsilon_4=1$ and $\upsilon_6=1$ Vibrationally Excited States: An Example of
  Strong Coriolis Interaction}},
\newblock \bibinfo{journal}{J. Mol. Spectrosc.} \bibinfo{volume}{216}
  (\bibinfo{year}{2002}) \bibinfo{pages}{335 -- 344}.
  \DOIprefix\doi{10.1006/jmsp.2002.8629}.
\bibitem[{Drouin(2017)}]{Drouin_Spfit_overview}
\bibinfo{author}{B.~J. Drouin},
\newblock \bibinfo{title}{{Practical uses of SPFIT}},
\newblock \bibinfo{journal}{J. Mol. Spectrosc.} \bibinfo{volume}{340}
  (\bibinfo{year}{2017}) \bibinfo{pages}{1 -- 15}.
  \DOIprefix\doi{10.1016/j.jms.2017.07.009}.
\bibitem[{Pickett(1991)}]{PICKETT1991}
\bibinfo{author}{H.~M. Pickett},
\newblock \bibinfo{title}{{The fitting and prediction of vibration-rotation
  spectra with spin interactions}},
\newblock \bibinfo{journal}{J. Mol. Spectrosc.} \bibinfo{volume}{148}
  (\bibinfo{year}{1991}) \bibinfo{pages}{371 -- 377}.
  \DOIprefix\doi{10.1016/0022-2852(91)90393-O}.
\bibitem[{Pickett(1972)}]{RAS_Coriolis}
\bibinfo{author}{H.~M. Pickett},
\newblock \bibinfo{title}{{Vibration—Rotation Interactions and the Choice of
  Rotating Axes for Polyatomic Molecules}},
\newblock \bibinfo{journal}{J. Chem. Phys.} \bibinfo{volume}{56}
  (\bibinfo{year}{1972}) \bibinfo{pages}{1715--1723}.
  \DOIprefix\doi{10.1063/1.1677430}.
\bibitem[{{Read} et~al.(1986){Read}, {Cohen}, and
  {Pickett}}]{cyanamide_rot_1986}
\bibinfo{author}{W.~G. {Read}}, \bibinfo{author}{E.~A. {Cohen}},
  \bibinfo{author}{H.~M. {Pickett}},
\newblock \bibinfo{title}{{The rotation-inversion spectrum of cyanamide}},
\newblock \bibinfo{journal}{J. Mol. Spectrosc.} \bibinfo{volume}{115}
  (\bibinfo{year}{1986}) \bibinfo{pages}{316--332}.
  \DOIprefix\doi{10.1016/0022-2852(86)90050-0}.
\bibitem[{{Kra{\'s}nicki} et~al.(2011){Kra{\'s}nicki}, {Kisiel}, {Jabs},
  {Winnewisser}, and {Winnewisser}}]{cyanamide_isos_rot_2011}
\bibinfo{author}{A.~{Kra{\'s}nicki}}, \bibinfo{author}{Z.~{Kisiel}},
  \bibinfo{author}{W.~{Jabs}}, \bibinfo{author}{B.~P. {Winnewisser}},
  \bibinfo{author}{M.~{Winnewisser}},
\newblock \bibinfo{title}{{Analysis of the mm- and submm-wave rotational
  spectra of isotopic cyanamide: New isotopologues and molecular geometry}},
\newblock \bibinfo{journal}{J. Mol. Spectrosc.} \bibinfo{volume}{267}
  (\bibinfo{year}{2011}) \bibinfo{pages}{144--149}.
  \DOIprefix\doi{10.1016/j.jms.2011.03.005}.
\bibitem[{{Coutens} et~al.(2019){Coutens}, {Zakharenko}, {Lewen},
  {J{\o}rgensen}, {Schlemmer}, and {M{\"u}ller}}]{cyanamide_w_15N_13C_rot_2019}
\bibinfo{author}{A.~{Coutens}}, \bibinfo{author}{O.~{Zakharenko}},
  \bibinfo{author}{F.~{Lewen}}, \bibinfo{author}{J.~K. {J{\o}rgensen}},
  \bibinfo{author}{S.~{Schlemmer}}, \bibinfo{author}{H.~S.~P. {M{\"u}ller}},
\newblock \bibinfo{title}{{Laboratory spectroscopic study of the $^{15}$N
  isotopomers of cyanamide, H$_{2}$NCN, and a search for them toward IRAS
  16293-2422 B}},
\newblock \bibinfo{journal}{Astron. Astrophys.} \bibinfo{volume}{623}
  (\bibinfo{year}{2019}) \bibinfo{pages}{A93}.
  \DOIprefix\doi{10.1051/0004-6361/201834605}.
  \href{http://arxiv.org/abs/1901.00421}{{\tt arXiv:1901.00421}}.
\bibitem[{{Christen} and {M{\"u}ller}(2003)}]{ethanediol_rot_2003}
\bibinfo{author}{D.~{Christen}}, \bibinfo{author}{H.~S.~P. {M{\"u}ller}},
\newblock \bibinfo{title}{{The millimeter wave spectrum of aGg ethylene glycol:
  The quest for higher precision}},
\newblock \bibinfo{journal}{Phys. Chem. Chem. Phys.} \bibinfo{volume}{5}
  (\bibinfo{year}{2003}) \bibinfo{pages}{3600}.
  \DOIprefix\doi{10.1039/B304566H}.
\bibitem[{{M{\"u}ller} and {Christen}(2004)}]{ethanediol_rot_2004}
\bibinfo{author}{H.~S.~P. {M{\"u}ller}}, \bibinfo{author}{D.~{Christen}},
\newblock \bibinfo{title}{{Millimeter and submillimeter wave spectroscopic
  investigations into the rotation-tunneling spectrum of gGg$^{'}$ ethylene
  glycol, HOCH$_{2}$CH$_{2}$OH}},
\newblock \bibinfo{journal}{J. Mol. Spectrosc.} \bibinfo{volume}{228}
  (\bibinfo{year}{2004}) \bibinfo{pages}{298--307}.
  \DOIprefix\doi{10.1016/j.jms.2004.04.009}.
\bibitem[{{Melosso} et~al.(2020){Melosso}, {Dore}, {Tamassia}, {Brogan},
  {Hunter}, and {McGuire}}]{ethanediol_rot_2020}
\bibinfo{author}{M.~{Melosso}}, \bibinfo{author}{L.~{Dore}},
  \bibinfo{author}{F.~{Tamassia}}, \bibinfo{author}{C.~L. {Brogan}},
  \bibinfo{author}{T.~R. {Hunter}}, \bibinfo{author}{B.~A. {McGuire}},
\newblock \bibinfo{title}{{The Submillimeter Rotational Spectrum of Ethylene
  Glycol up to 890 GHz and Application to ALMA Band 10 Spectral Line Data of
  NGC 6334I}},
\newblock \bibinfo{journal}{J. Phys. Chem. A} \bibinfo{volume}{124}
  (\bibinfo{year}{2020}) \bibinfo{pages}{240--246}.
  \DOIprefix\doi{10.1021/acs.jpca.9b10803}.
  \href{http://arxiv.org/abs/1912.01472}{{\tt arXiv:1912.01472}}.
\bibitem[{{Pearson} and {Drouin}(2005)}]{propargyl__alcohol_rot_2005}
\bibinfo{author}{J.~C. {Pearson}}, \bibinfo{author}{B.~J. {Drouin}},
\newblock \bibinfo{title}{{The ground state torsion-rotation spectrum of
  propargyl alcohol (HCCCH$_{2}$OH)}},
\newblock \bibinfo{journal}{J. Mol. Spectrosc.} \bibinfo{volume}{234}
  (\bibinfo{year}{2005}) \bibinfo{pages}{149--156}.
  \DOIprefix\doi{10.1016/j.jms.2005.08.013}.
\bibitem[{{M{\"u}ller} et~al.(2010){M{\"u}ller}, {Dong}, {Nesbitt}, {Furuya},
  and {Saito}}]{H2DO+_analysis_rot_2010}
\bibinfo{author}{H.~S.~P. {M{\"u}ller}}, \bibinfo{author}{F.~{Dong}},
  \bibinfo{author}{D.~J. {Nesbitt}}, \bibinfo{author}{T.~{Furuya}},
  \bibinfo{author}{S.~{Saito}},
\newblock \bibinfo{title}{{Tunneling dynamics and spectroscopic parameters of
  monodeuterated hydronium, H$_2$DO$^+$, from a combined analysis of infrared
  and sub-millimeter spectra}},
\newblock \bibinfo{journal}{Phys. Chem. Chem. Phys.} \bibinfo{volume}{12}
  (\bibinfo{year}{2010}) \bibinfo{pages}{8362}.
  \DOIprefix\doi{10.1039/c002067b}.
\bibitem[{{M{\"u}ller} et~al.(2016){M{\"u}ller}, {Belloche}, {Xu}, {Lees},
  {Garrod}, {Walters}, {van Wijngaarden}, {Lewen}, {Schlemmer}, and
  {Menten}}]{ethanethiol_RAS-Fit_2016}
\bibinfo{author}{H.~S.~P. {M{\"u}ller}}, \bibinfo{author}{A.~{Belloche}},
  \bibinfo{author}{L.-H. {Xu}}, \bibinfo{author}{R.~M. {Lees}},
  \bibinfo{author}{R.~T. {Garrod}}, \bibinfo{author}{A.~{Walters}},
  \bibinfo{author}{J.~{van Wijngaarden}}, \bibinfo{author}{F.~{Lewen}},
  \bibinfo{author}{S.~{Schlemmer}}, \bibinfo{author}{K.~M. {Menten}},
\newblock \bibinfo{title}{{Exploring molecular complexity with ALMA (EMoCA):
  Alkanethiols and alkanols in Sagittarius B2(N2)}},
\newblock \bibinfo{journal}{Astron. Astrophys.} \bibinfo{volume}{587}
  (\bibinfo{year}{2016}) \bibinfo{pages}{A92}.
  \DOIprefix\doi{10.1051/0004-6361/201527470}.
  \href{http://arxiv.org/abs/1512.05301}{{\tt arXiv:1512.05301}}.
\bibitem[{{Margul{\`e}s} et~al.(2017){Margul{\`e}s}, {McGuire}, {Senent},
  {Motiyenko}, {Remijan}, and {Guillemin}}]{hydroxyacetonitrile_rot_2017}
\bibinfo{author}{L.~{Margul{\`e}s}}, \bibinfo{author}{B.~A. {McGuire}},
  \bibinfo{author}{M.~L. {Senent}}, \bibinfo{author}{R.~A. {Motiyenko}},
  \bibinfo{author}{A.~{Remijan}}, \bibinfo{author}{J.~C. {Guillemin}},
\newblock \bibinfo{title}{{Submillimeter spectra of 2-hydroxyacetonitrile
  (glycolonitrile; HOCH$_{2}$CN) and its searches in GBT PRIMOS observations of
  Sgr B2(N)}},
\newblock \bibinfo{journal}{Astron. Astrophys.} \bibinfo{volume}{601}
  (\bibinfo{year}{2017}) \bibinfo{pages}{A50}.
  \DOIprefix\doi{10.1051/0004-6361/201628551}.
\bibitem[{{Bermudez} et~al.(2017){Bermudez}, {Bailleux}, and
  {Cernicharo}}]{hydroxymethyl_rot_2017}
\bibinfo{author}{C.~{Bermudez}}, \bibinfo{author}{S.~{Bailleux}},
  \bibinfo{author}{J.~{Cernicharo}},
\newblock \bibinfo{title}{{Laboratory detection of the rotational-tunnelling
  spectrum of the hydroxymethyl radical, CH$_{2}$OH}},
\newblock \bibinfo{journal}{Astron. Astrophys.} \bibinfo{volume}{598}
  (\bibinfo{year}{2017}) \bibinfo{pages}{A9}.
  \DOIprefix\doi{10.1051/0004-6361/201629508}.
\bibitem[{{Chitarra} et~al.(2020){Chitarra}, {Martin-Drumel}, {Gans}, {Loison},
  {Spezzano}, {Lattanzi}, {M{\"u}ller}, and {Pirali}}]{hydroxymethyl_rot_2020}
\bibinfo{author}{O.~{Chitarra}}, \bibinfo{author}{M.-A. {Martin-Drumel}},
  \bibinfo{author}{B.~{Gans}}, \bibinfo{author}{J.-C. {Loison}},
  \bibinfo{author}{S.~{Spezzano}}, \bibinfo{author}{V.~{Lattanzi}},
  \bibinfo{author}{H.~S.~P. {M{\"u}ller}}, \bibinfo{author}{O.~{Pirali}},
\newblock \bibinfo{title}{{Reinvestigation of the rotation-tunneling spectrum
  of the CH$_{2}$OH radical. Accurate frequency determination of transitions of
  astrophysical interest up to 330 GHz}},
\newblock \bibinfo{journal}{Astron. Astrophys.} \bibinfo{volume}{644}
  (\bibinfo{year}{2020}) \bibinfo{pages}{A123}.
  \DOIprefix\doi{10.1051/0004-6361/202039071}.
\bibitem[{Herschbach and Swalen(1958)}]{Herschbach_PO_forbidden_E_lines}
\bibinfo{author}{D.~R. Herschbach}, \bibinfo{author}{J.~D. Swalen},
\newblock \bibinfo{title}{Internal barrier of propylene oxide from the
  microwave spectrum. ii},
\newblock \bibinfo{journal}{J. Chem. Phys.} \bibinfo{volume}{29}
  (\bibinfo{year}{1958}) \bibinfo{pages}{761--776}.
  \DOIprefix\doi{10.1063/1.1744588}.
\bibitem[{Wehres et~al.(2017)Wehres, Heyne, Lewen, Hermanns, Schmidt, Endres,
  Graf, Higgins, and Schlemmer}]{wehres2017_emission}
\bibinfo{author}{N.~Wehres}, \bibinfo{author}{B.~Heyne},
  \bibinfo{author}{F.~Lewen}, \bibinfo{author}{M.~Hermanns},
  \bibinfo{author}{B.~Schmidt}, \bibinfo{author}{C.~Endres},
  \bibinfo{author}{U.~U. Graf}, \bibinfo{author}{D.~R. Higgins},
  \bibinfo{author}{S.~Schlemmer},
\newblock \bibinfo{title}{100 {GH}z room-temperature laboratory emission
  spectrometer},
\newblock \bibinfo{journal}{Proc. Int. Astron. Union} \bibinfo{volume}{13}
  (\bibinfo{year}{2017}) \bibinfo{pages}{332–345}.
  \DOIprefix\doi{10.1017/S1743921317007803}.
\bibitem[{Wehres et~al.(2018)Wehres, Maßen, Borisov, Schmidt, Lewen, Graf,
  Honingh, Higgins, and Schlemmer}]{wehres2018_emission}
\bibinfo{author}{N.~Wehres}, \bibinfo{author}{J.~Maßen},
  \bibinfo{author}{K.~Borisov}, \bibinfo{author}{B.~Schmidt},
  \bibinfo{author}{F.~Lewen}, \bibinfo{author}{U.~U. Graf},
  \bibinfo{author}{C.~E. Honingh}, \bibinfo{author}{D.~R. Higgins},
  \bibinfo{author}{S.~Schlemmer},
\newblock \bibinfo{title}{A laboratory heterodyne emission spectrometer at
  submillimeter wavelengths},
\newblock \bibinfo{journal}{Phys. Chem. Chem. Phys.} \bibinfo{volume}{20}
  (\bibinfo{year}{2018}) \bibinfo{pages}{5530--5544}.
  \DOIprefix\doi{10.1039/C7CP06394F}.

\end{thebibliography}

\newpage
\section*{Supplementary material}

\setcounter{figure}{0}
\renewcommand{\thefigure}{A\arabic{figure}}

\begin{figure*}[t]
\centering
\includegraphics[width=0.9\linewidth]{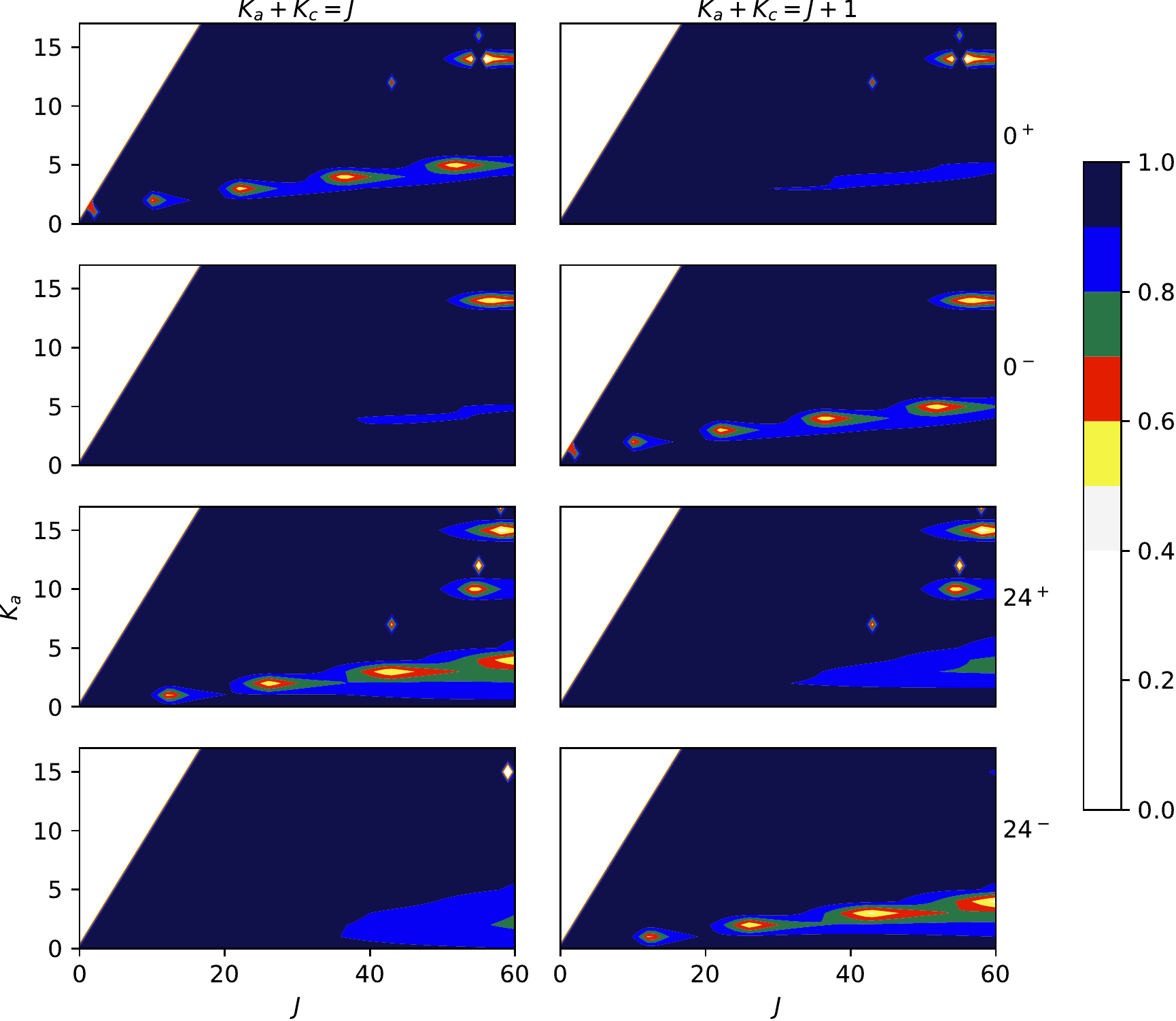}
\caption{
Contour plots of mixing coefficients of \textit{gauche}-propanal of energy levels with $K_a+K_c=J$ and $K_a+K_c=J+1$ on the left- and right-hand side, respectively. The studied tunneling states $0^+$, $0^-$, $24^+$, and $24^-$ are plotted from top to bottom.
Tunneling-rotation interaction are prominently seen for energy levels of $0^+$ ($K_a+K_c=J$) and $0^-$ ($K_a+K_c=J+1$) with $K_a=1,2,3,4,5$ around $J=2,10,22,36,52$, respectively, as well as of $24^+$ ($K_a+K_c=J$) and $24^-$ ($K_a+K_c=J+1$) with $K_a=1,2,3$ around $J=12,26,42$, respectively.
Furthermore, $\Delta K_a=5$ interaction between $0^+$ ($K_a=12$) and $24^+$ ($K_a=7$) are locally seen around $J=43$, the $\Delta K_a=4$ interaction around $J=55$ are seen for $0^+$ as well as $0^-$ at $K_a=14$ and $24^+$ at $K_a=10$. Label switching is occurring at $K_a=14$ of $0^+$ which seem to divide the interaction area. This can also be seen by two $24^+$ (green markers) energy levels and a missing blue one at $J=55$ in Fig.~\ref{Fig:8_Red-Egy-interactions}b of the main article. 
Additionally, strong mixing is noticed for $K_a=15$ of $24^+$, where an interaction with $K_a=18$ of $\upsilon=0$ is predicted. 
}
\label{FigA:1_mixing}
\end{figure*}

\begin{figure*}[t]
\centering
\includegraphics[width=0.49\linewidth]{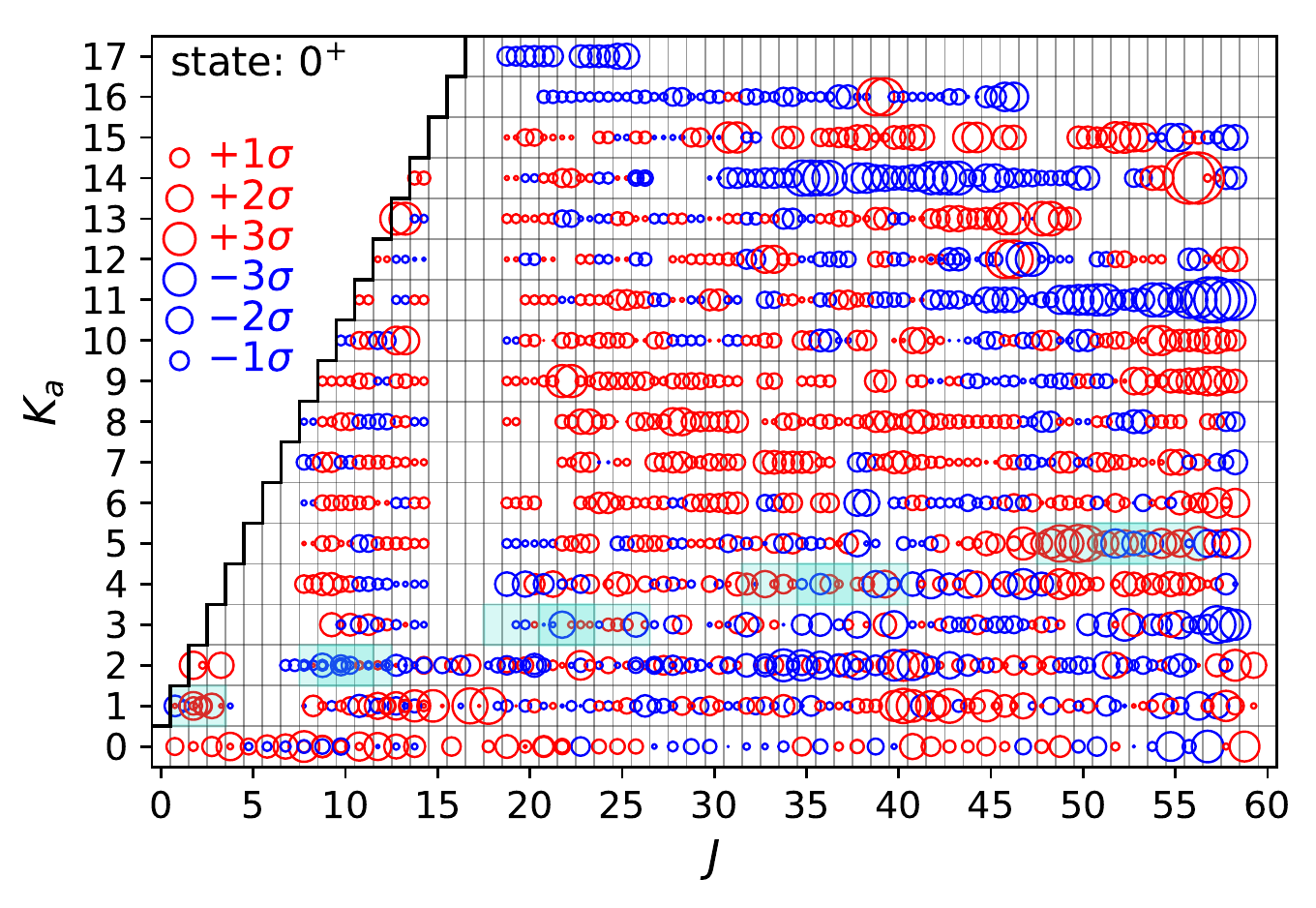}
\includegraphics[width=0.49\linewidth]{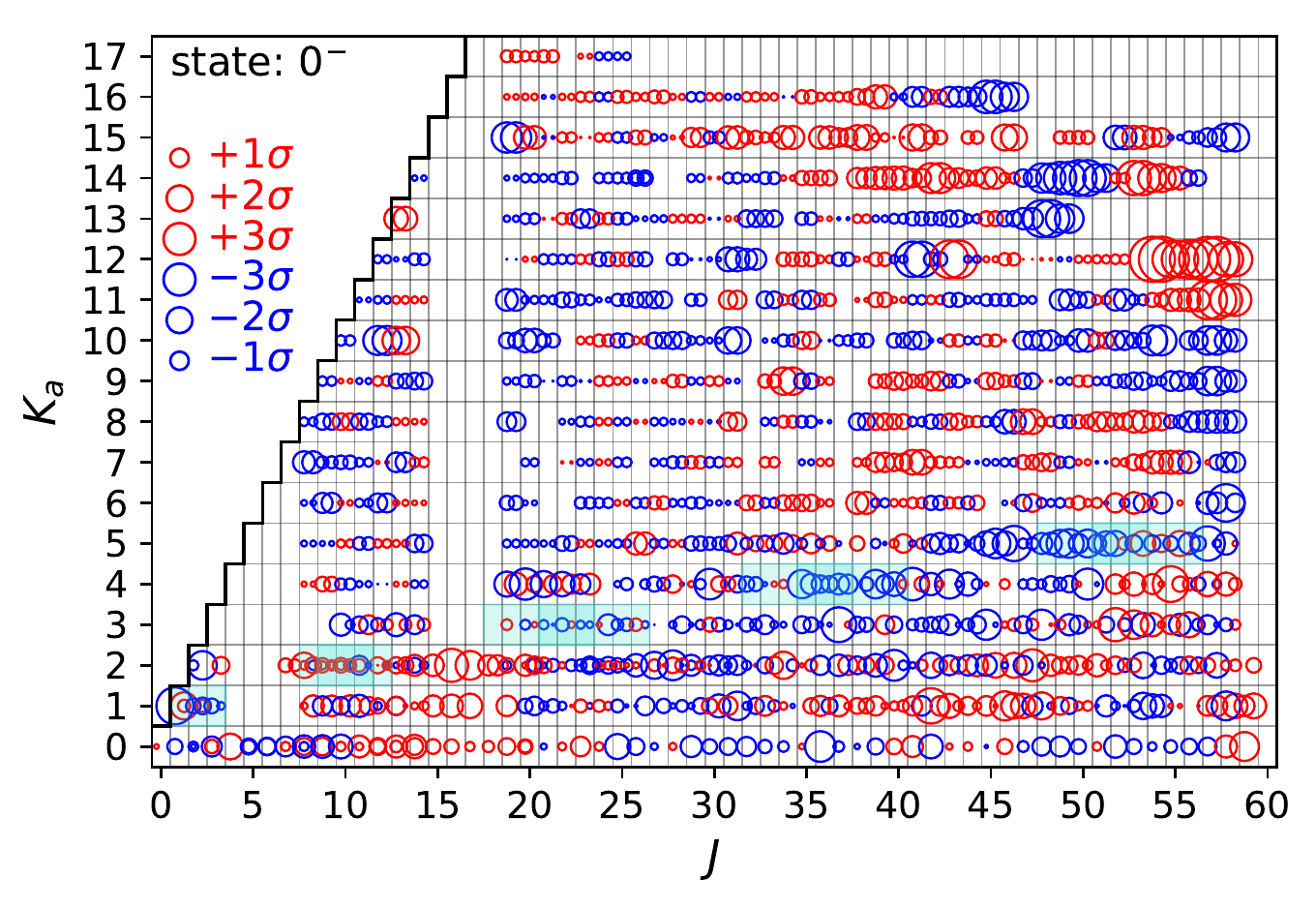}
\includegraphics[width=0.49\linewidth]{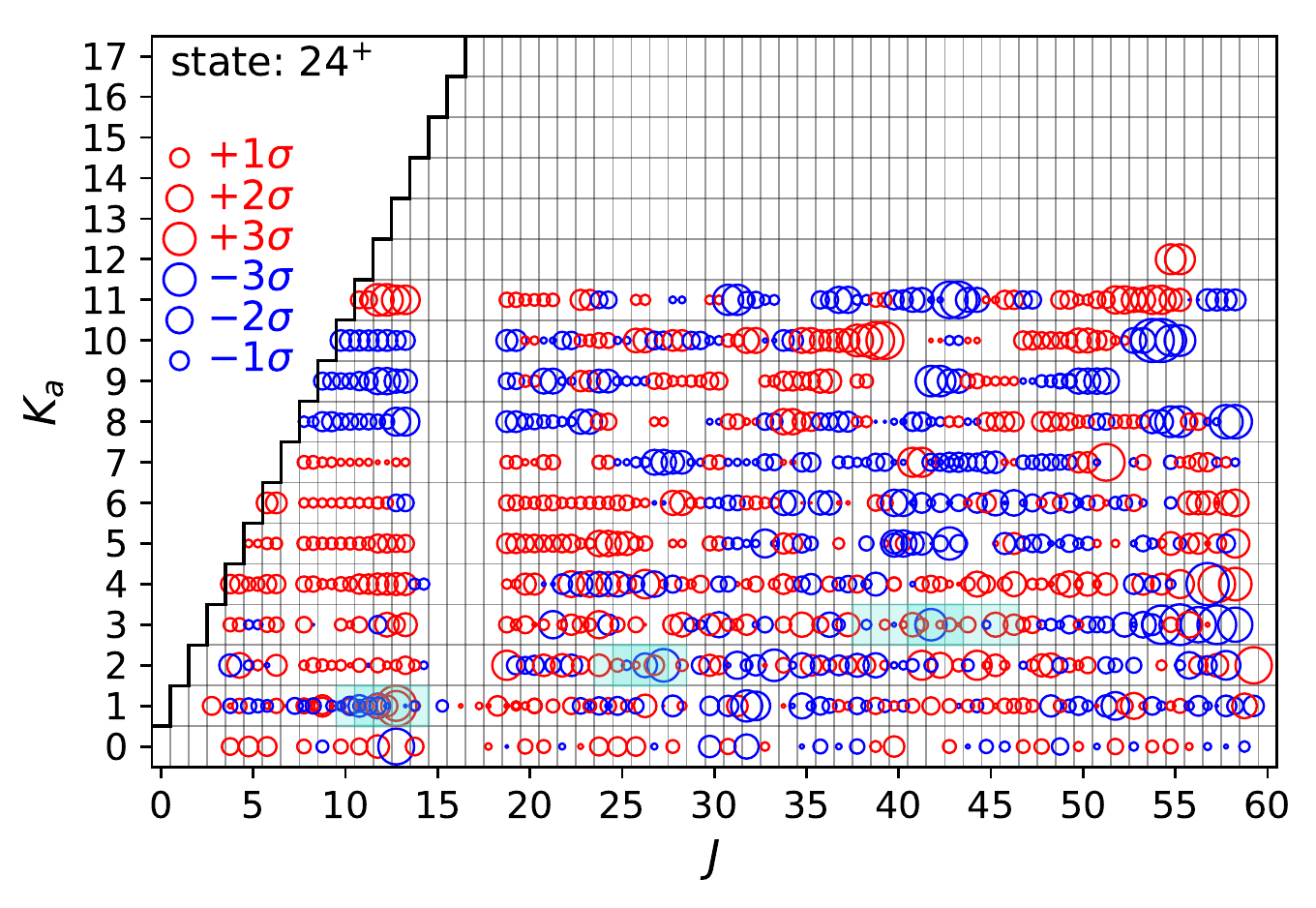}
\includegraphics[width=0.49\linewidth]{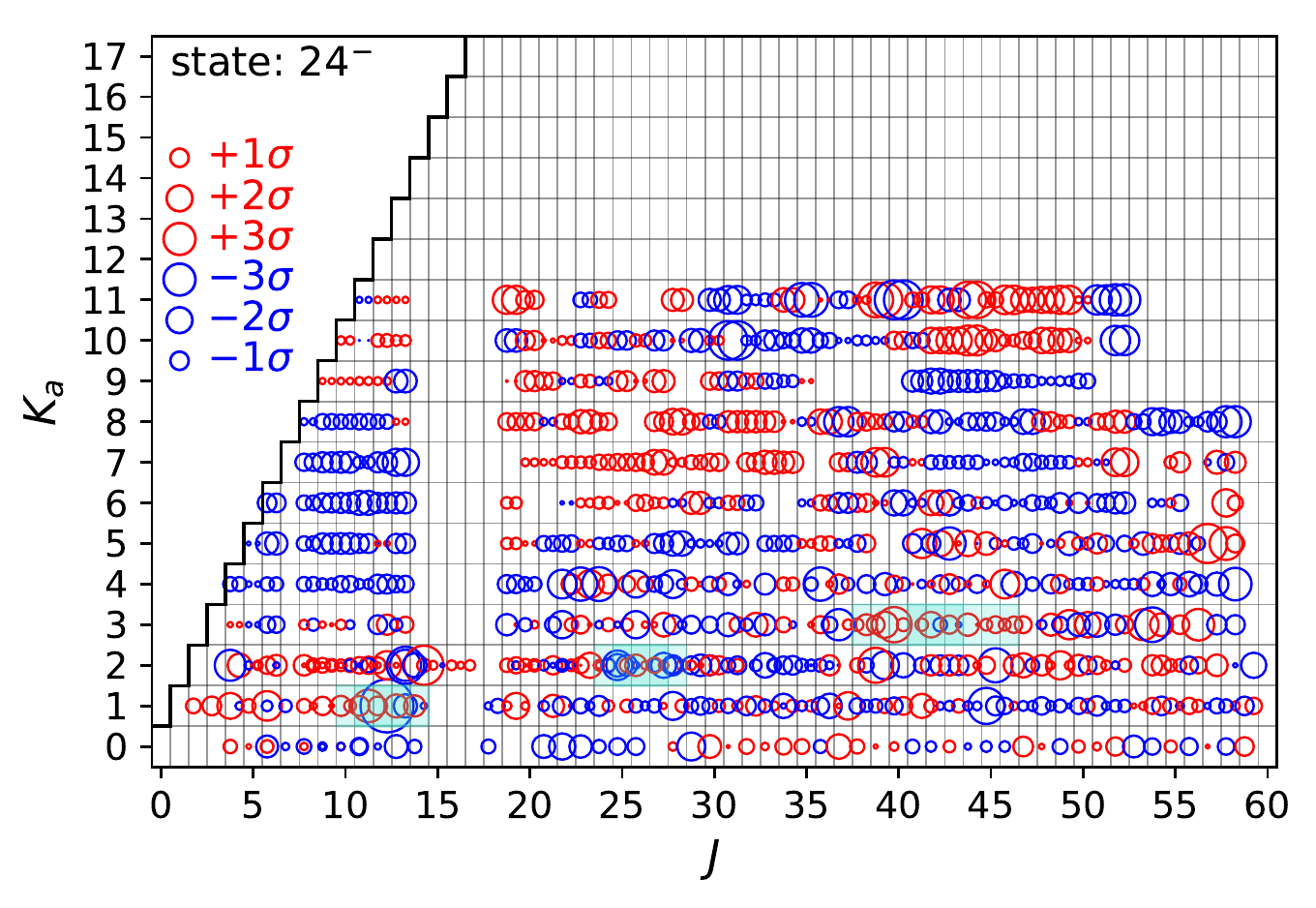}
\caption{
Quantum number coverage of fit transitions of the four tunneling states, $0^+$, $0^-$, $24^+$, and  $24^-$, of \textit{gauche}-propanal. The size of the marker cycles is proportional to the deviations between observed and predicted center frequencies $(\nu_{obs.}-\nu_{calc.})/\Delta \nu$, based on the spectroscopic parameters in Table~\ref{Tab:Spectroscopic_parameters}.  
}
\label{FigA:2_QN}
\end{figure*}

\begin{figure}[t]
\centering
\includegraphics[width=1.0\linewidth]{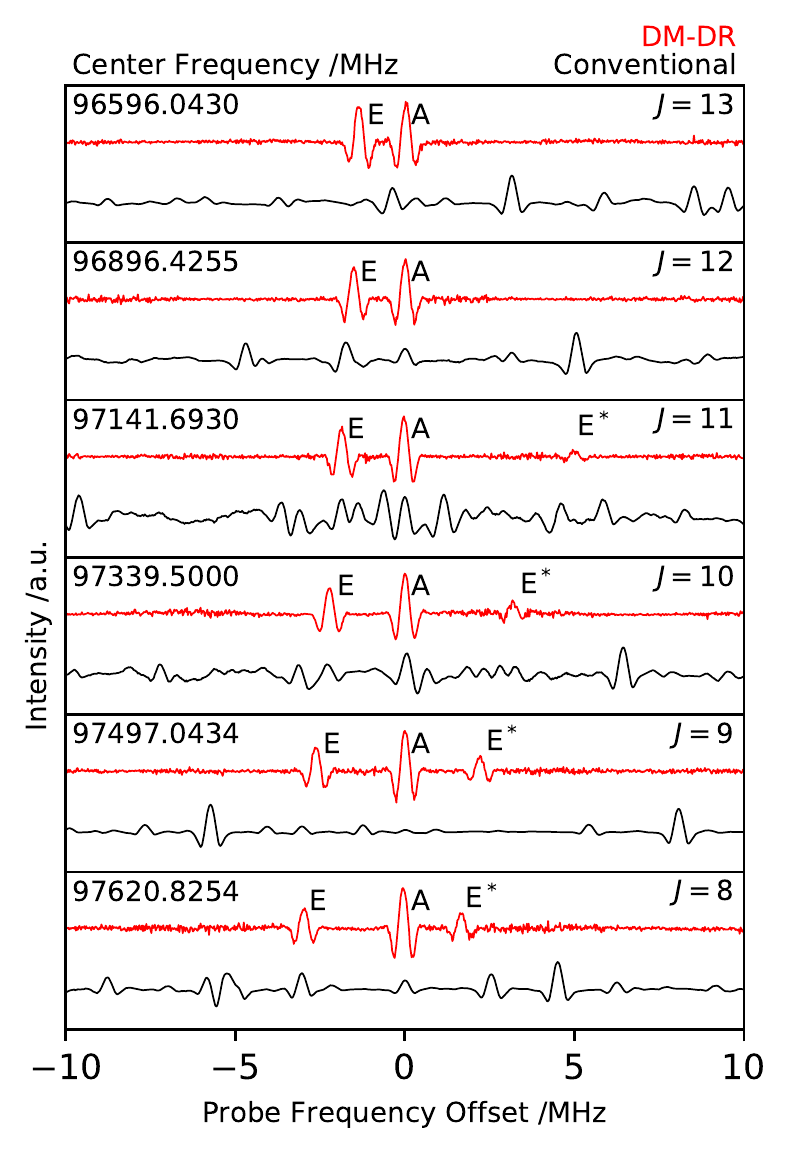}
\caption{DM-DR measurements (in red) of $Q$-branch $c$-type transitions which connect the two tunneling states ($J_{K_a,K_c}=J_{3,J-2}\leftarrow J_{2,J-2}$; $24^+\leftarrow24^-$). Assignments of these transitions is hampered in conventional spectra (in black). These transitions show methyl internal rotation splitting (A, E) and can be unambiguously assigned due to nominally forbidden E$^*$ transitions in rigid rotors, which share their intensity with E ones \cite{Herschbach_PO_forbidden_E_lines}.}
\label{FigA:3_c-types_unp}
\end{figure}

\begin{figure*}[t]
\centering
\includegraphics[width=0.9\linewidth]{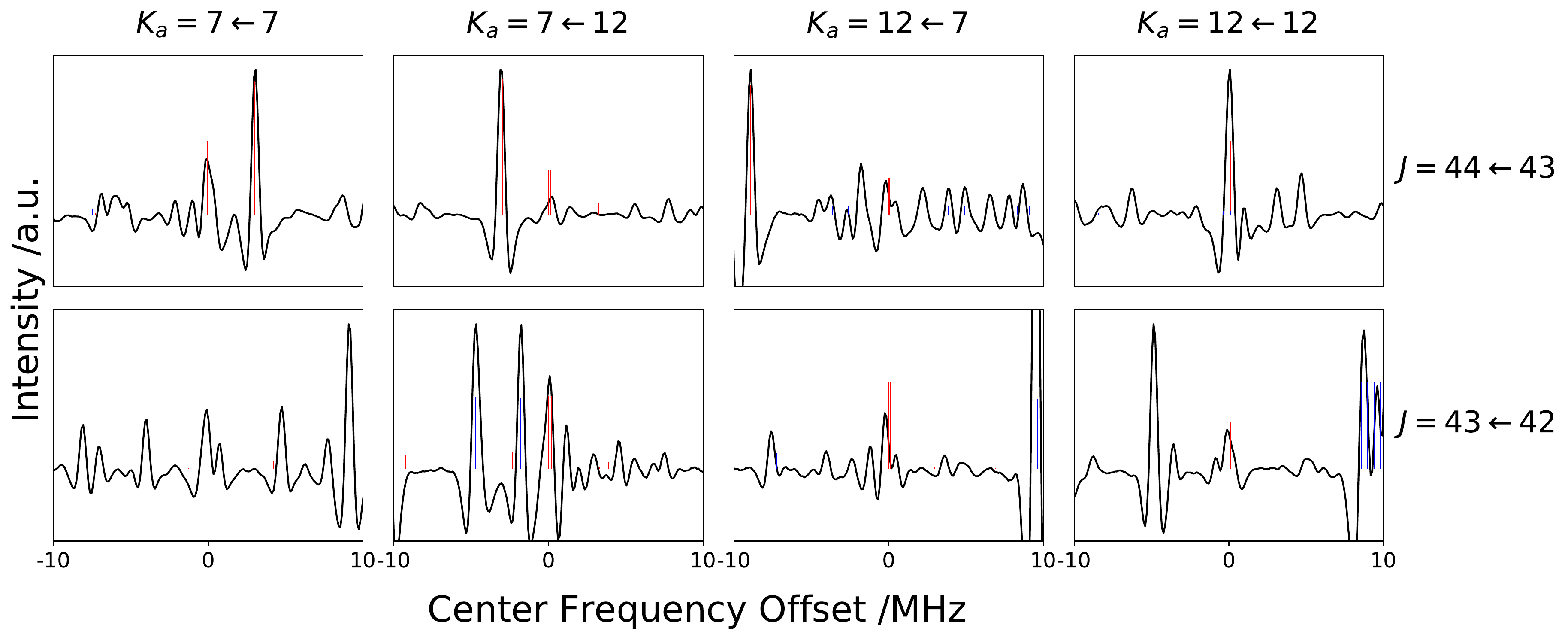}
\caption{
Strongly perturbed transitions of the Fermi resonance ($\Delta K_a=5$ interaction) between $0^+$ ($K_a=12$) and $24^+$ ($K_a=7$) around $J=43$ of \textit{gauche}-propanal.
The experimental spectrum is shown in black and predictions of \textit{syn}- and \textit{gauche}-propanal are shown as stick spectra in blue and red, respectively.
Mixing of energy levels of $0^+$ and $24^+$ of \textit{gauche}-propanal, cf. Fig.~\ref{Fig:8_Red-Egy-interactions}b, allow to observe four nominally forbidden transitions (with $K_a=12\leftrightarrow 7$ and $0^+\leftrightarrow24^+$), in addition to the four nominally allowed $^qR$ series transitions (with either $K_a=12\leftarrow 12$ and $0^+\leftarrow0^+$ or $K_a=7\leftarrow 7$ and $24^+\leftarrow24^+$).
The eight transitions are marked by the eight arrows in the reduced energy diagram in Fig.~\ref{Fig:8_Red-Egy-interactions}b and all of them show an internal structure of their line profile.
The methyl group internal rotation is not treated in this analysis as no splitting is observed for a substantial majority of lines, in particular for $^qR$ series transitions, see Sec.~\ref{SubSec:v24} for more information.
However, the conspicuous line shapes are considered to be an additional conformation of the correct assignments of the eight strongly perturbed lines.
The internal structure is expected to originate from a mixing of different methyl internal rotation splittings of the two vibrational states.
In the final analysis, weighted experimental intensities of the internal structure are used to derive center frequencies and these are added to the line list with somewhat larger uncertainties (100\,kHz) compared to other frequencies to account for the untreated internal structure. DR experiments may be another tool to verify the assignments in the future, if powerful sources are available at these high frequencies.
}
\label{FigA:4_Ka5_transitions}
\end{figure*}

The appearance of tunneling-rotation interaction, Fermi resonance and Coriolis interaction is visualized by plotting the mixing coefficients of the energy levels in dependence of $K_a$ and $J$ quantum numbers, see Fig.~\ref{FigA:1_mixing}. The quantum number coverage of assigned and fit transitions is depicted in Figs.~\ref{FigA:2_QN}.
Additionally, the experimental assignments of methyl internal rotation components (A, E, and E$^*$) is demonstrated by the appearance of nominally forbidden E$^*$ transitions, which share their intensity with the E transitions \cite{Herschbach_PO_forbidden_E_lines}, with $Q$-branch $c$-type transitions $J_{K_a,K_c}=J_{3,J-2}\leftarrow J_{2,J-2}$ in Fig.~\ref{FigA:3_c-types_unp}. 
Furthermore, the transitions resulting from most perturbed energy levels of the $\Delta K_a=5$ interaction are shown in Fig.~\ref{FigA:4_Ka5_transitions}.
The results of the millimeter-wave relative intensity measurements are summarized in Table~\ref{TabA:MMW_Rel_Intensity_v24}.

internal structure weighed by experimental relative intensity

\begin{table*}[t]
\centering
\caption{Millimeter-wave relative intensity measurements of $\upsilon=0$ and $\upsilon_{24}=1$ of \textit{gauche}-propanal.}
\begin{tabular}{ r  r r r r r r r}
\hline
\multicolumn{1}{l}{Transition}   & Frequency$^a$     & \multicolumn{4}{c}{Intensity$^b$ /a.u.} & \multicolumn{1}{c}{$R^c$} & \multicolumn{1}{c}{$\overline{R}$} \\
                   &   \multicolumn{1}{c}{/MHz} & $0^+$ & $0^-$ & $24^+$ & $24^-$ \\
\hline
\multicolumn{1}{l}{$J_{0,~J}\leftarrow (J-1)_{0,~J-1}$}     \\
 $J=10$ &  ~~~~~84433.3080 & ~~~27555    &  ~~~27006   & ~~~20145    & ~~~23525   & ~~~0.80  \\
 $J=13$ & 109592.1802 & 12673    &  11678   & 9227     & 10849   &  0.82 \\
 $J=14$ & 117952.3133 & 5045     &  4301    & 4055     & 4413    & 0.91  \\
 $J=22$ & 184344.5475 & 185815   &  180231  & 127299   & 146387  & 0.75  \\
 $J=33$ & 274651.7827 & 49280    &  42338   & 32034    & 29777   & 0.67  \\
 $J=49$ & 404836.5234 & 5251     &  7002    & 4720     & 2745    & 0.61 & ~~~0.76(11) \\
~\\
\multicolumn{1}{l}{$11_{K_a, K_c}\leftarrow 10_{K_a, K_c-1}$} \\
 $K_a=4$ & 93136.0761 & 31346    &  35416   & 31820    & 27102   &  0.88 \\
 $K_a=5$ & 93165.8352 & 33457    &  35058   & 21810    & 25410   &  0.69 \\
 $K_a=6$ & 93206.2575 & 28107    &  32080   & 19931    & 20957   &  0.68 \\
 $K_a=7$ & 93255.8588 & 26711    &  23473   & 18544    & 16524   &  0.70 \\
 $K_a=8$ & 93314.0764 & 20306    &  22892   & 18017    & 14072   &  0.74 \\
 $K_a=9$ & 93380.6683 & 15614    &  16589   & 16244    & 12175   &  0.88 &  0.76(9)\\
~\\
         &          &          &          &         &  & & \textbf{0.76(9)}\\
\hline
\end{tabular}
\begin{footnotesize}
\begin{flushleft}
    \vspace{0.5em}
    {$^a$ The frequency is given for the $0^+$ state.\\
  $^b$  The intensity is derived by the difference of maximum and minimum intensities of a measured line.\\
  $^c$ The ratio $R$ is simply calculated by considering all Intensities $I_i$ with $R=(I_{24^+}+I_{24^-})/(I_{0^+}+I_{0^-})$. All considered transitions are $a$-type transitions and differences in line shapes as well as frequencies are neglected as transitions within one row appear in a small frequency window, in particular $(\mu^2_a\nu^2S_{Doppler})$ is assumed to be equal for transitions in the same row.
}
\end{flushleft}
\end{footnotesize}
\label{TabA:MMW_Rel_Intensity_v24}
\end{table*}

\end{document}